\documentclass[acmsmall,screen]{acmart}

\usepackage{fancyvrb}
\usepackage{xurl}
\usepackage{booktabs}
\usepackage{url}
\usepackage{amsmath,amsfonts}
\usepackage{algorithmic}
\usepackage{graphicx}
\usepackage{textcomp}
\usepackage{xspace}
\usepackage{subcaption}
\usepackage{tabularx}
\usepackage{framed}
\usepackage{chngpage}
\usepackage[many]{tcolorbox}
\usepackage{pgfplots}
\pgfplotsset{compat=newest}
\usepackage{siunitx}
\usepackage{balance}
\usepackage{enumitem}
\usepackage{sepnum}
\usepackage{multirow}
\usepackage[htt]{hyphenat}

\usepackage{hyperref}

\usepackage{tikz}
\usepackage[group-separator={,}, group-minimum-digits=4]{siunitx}

\usepackage{pgf-pie}
\pgfplotsset{compat=1.18}

\newcommand{\refactoring}[1]{{\sc\small #1}}

\newcommand{\rqa}{$RQ_1$}
\newcommand{\rqb}{$RQ_2$}
\newcommand{\rqc}{$RQ_3$}
\newcommand{\rqd}{$RQ_4$}
\newcommand{\rqaa}{How prevalent is refactoring in agentic commits?}
\newcommand{\rqbb}{What are the common types of agentic refactoring?}
\newcommand{\rqcc}{What is the purpose of agentic refactoring?}
\newcommand{\rqdd}{To what extent does agentic refactoring affect code quality?}

\newcommand{\rqA}{\rqa: \rqaa}
\newcommand{\rqB}{\rqb: \rqbb}
\newcommand{\rqC}{\rqc: \rqcc}
\newcommand{\rqD}{\rqd: \rqdd}

\newcommand{\AR}{Agentic Refactoring\xspace}

\newcommand{\RMiner}{\texttt{RefactoringMiner}\xspace}  
\newcommand{\designitejava}{\texttt{DesigniteJava}\xspace}

\newcommand{\numint}[1]{\num[round-mode=places, round-precision=0, group-digits=true]{#1}}

\newcommand{\codex}{OpenAI Codex\xspace}
\newcommand{\devin}{Devin\xspace}
\newcommand{\cursor}{Cursor\xspace}
\newcommand{\claude}{Claude Code\xspace}

\newcommand{\TotalCommits}{14998}
\newcommand{\TotalPRs}{12256}

\newcommand{\ReposBeforeFiltering}{3232}
\newcommand{\TotalRepos}{1613}

\newcommand{\toyProjectNum}{1235}
\newcommand{\uncertainProjectNum}{362}

\newcommand{\NonToyProjectNum}{1635}

\newcommand{\ClosedPRs}{11504}
\newcommand{\MergedPRs}{10645}
\pgfmathsetmacro{\UnMergedPRsResult}{\ClosedPRs-\MergedPRs}

\newcommand{\ClosedPRsPct}{93.9\%\xspace}

\newcommand{\CodexCommits}{13389}
\newcommand{\DevinCommits}{860}
\newcommand{\CursorCommits}{663}
\newcommand{\ClaudeCommits}{86}

\newcommand{\CodexCommitPct}{89.3\%\xspace}
\newcommand{\DevinCommitPct}{5.7\%\xspace}
\newcommand{\CursorCommitPct}{4.4\%\xspace}
\newcommand{\ClaudeCommitPct}{0.6\%\xspace}

\newcommand{\CodexPRs}{11557}
\newcommand{\DevinPRs}{335}
\newcommand{\CursorPRs}{337}
\newcommand{\ClaudePRs}{27}

\newcommand{\CodexPRPct}{94.3\%\xspace}
\newcommand{\DevinPRPct}{2.7\%\xspace}
\newcommand{\CursorPRPct}{2.7\%\xspace}
\newcommand{\ClaudePRPct}{0.2\%\xspace}

\newcommand{\TotalAgenticCommits}{\TotalCommits}
\newcommand{\RefCommits}{5789}

\newcommand{\AgenticRefCommits}{3907}

\newcommand{\AgentRefCommitPct}{26.1}
\newcommand{\AgenticRefInstances}{51412}

\pgfmathsetmacro{\NonSARCommits}{1882}
\pgfmathsetmacro{\NonSARCommitsPct}{32.5}
\pgfmathsetmacro{\NonSARInstances}{18801}
\pgfmathsetmacro{\NonSARInstancesPct}{36.6}

\newcommand{\OthersCommits}{11091}
\newcommand{\OthersCommitsPct}{73.9}

\newcommand{\SARCommits}{3907}

\newcommand{\SARInstances}{32611}
\newcommand{\SARInstancePct}{63.4}

\newcommand{\ARMaintainabilityPct}{52.5\%}
\newcommand{\ARReadabilityPct}{28.1\%}
\newcommand{\ARDuplicationPct}{1.1\%}
\newcommand{\ARReusePct}{4.6\%}

\newcommand{\ARDebugyPct}{1.9\%}
\newcommand{\ARLagacyPct}{2.0\%}

\newcommand{\AgenticHighPct}{52.4\%}
\newcommand{\AgenticMediumPct}{20.0\%}
\newcommand{\AgenticLowPct}{27.6\%}
\newcommand{\HumanHighPct}{54.9\%}
\newcommand{\HumanMediumPct}{20.7\%}
\newcommand{\HumanLowPct}{24.4\%}

\newcommand{\AgentMoveClassPct}{5.3\%}
\newcommand{\AgentRenameAttributePct}{5.3\%}
\newcommand{\AgentRenameMethodPct}{4.3\%}

\newcommand{\AgentChangeAttributeTypePct}{4.6\%}
\newcommand{\AgentChangeParameterTypePct}{3.6\%}
\newcommand{\AgentExtractMethodPct}{3.0\%}

\newcommand{\AgentRenameVariablePct}{7.6\%}
\newcommand{\AgentChangeVariableTypePct}{6.8\%}
\newcommand{\AgentRenameParameterPct}{4.9\%}

\newcommand{\AgentTopLowSumPct}{19.3\%}   
\newcommand{\HumanAddMethodAnnotationPct}{5.8\%}
\newcommand{\HumanRenameMethodPct}{4.4\%}
\newcommand{\HumanAddParameterPct}{4.3\%}

\newcommand{\HumanChangeParameterTypePct}{6.2\%}
\newcommand{\HumanChangeReturnTypePct}{4.7\%}
\newcommand{\HumanChangeAttributeTypePct}{3.5\%}

\newcommand{\HumanChangeVariableTypePct}{7.7\%}
\newcommand{\HumanRenameVariablePct}{3.3\%}
\newcommand{\HumanRenameParameterPct}{3.0\%}

\newcommand{\HumanTopLowSumPct}{14.0\%}   

\newcommand{\PurposeHumanKappa}{0.83}
\newcommand{\PurposeHumanAccuracy}{0.85}
\newcommand{\PurposeHumanMacroF}{0.85}
\newcommand{\PurposeGPTKappa}{0.77}
\newcommand{\PurposeGPTAccuracy}{0.80}
\newcommand{\PurposeGPTMacroF}{0.80}
\newcommand{\PurposeReliabilityN}{79}
\newcommand{\PurposeGPTDisagree}{16}
\newcommand{\PurposeHumanDisagree}{12}
\newcommand{\ARLogicalMismatch}{446}
\newcommand{\ARLogicalMismatchPct}{11.4}
\newcommand{\ARMotivationBase}{3461}

\newcommand{\GPTCostTotal}{\$4}
\newcommand{\GPTCostTokens}{5.58 million}

\newcommand{\ValidationSampleTotal}{344}
\newcommand{\ValidationSampleExcluded}{25}
\newcommand{\ValidationSampleAnalyzed}{319}

\newcommand{\ValidationAnnotatorOneExp}{8}
\newcommand{\ValidationAnnotatorTwoExp}{9}
\newcommand{\ValidationAnnotatorThreeExp}{18}

\newcommand{\ValidationHumanKappa}{0.83}
\newcommand{\ValidationHumanAgreement}{0.92}
\newcommand{\ValidationHumanMacroF}{0.92}
\newcommand{\ValidationGPTKappa}{0.80}
\newcommand{\ValidationGPTAgreement}{0.90}
\newcommand{\ValidationGPTMacroF}{0.90}

\definecolor{darkgreen}{rgb}{0, 0.5, 0} 
\definecolor{whitesmoke}{rgb}{0.99, 0.99, 0.99} 

\def\sec#1{Section~\ref{#1}}
\def\Underline{\setbox0\hbox\bgroup\let\\\endUnderline}
\def\endUnderline{\vphantom{y}\egroup\smash{\underline{\box0}}\\}
\def\|{\verb|}

\newcommand{\ie}{\textit{i.e.,}\xspace}
\newcommand{\eg}{\textit{e.g.,}\xspace}

\newcommand{\etc}{\xspace\textit{etc.}\xspace}
\newcommand{\etal}{\xspace\textit{et al.}\xspace}

\newcounter{findingctr}

\newcommand{\finding}[1]{%
  \refstepcounter{findingctr}%
  \textbf{Finding~\#\thefindingctr: #1}%
}

\usepackage{listings}
\definecolor{backcolour}{rgb}{0.95,0.95,0.92}
\lstdefinelanguage{diff}{
  morecomment=**[f][\color{red}]{-},         
  morecomment=**[f][\color{darkgreen}]{+},       
  moredelim=**[is][\bfseries]{@@}{@@},
}
\definecolor{backcolour}{rgb}{0.95,0.95,0.92}
\lstdefinelanguage{commit}{ 
  breakindent = 0pt,
  numbers=none,
  backgroundcolor=\color{white},
  frame=single,
  xleftmargin=3.5em,
  numbersep=0em,
  xrightmargin=1.5em,
}

\usepackage[many]{tcolorbox}  
\tcbuselibrary{listings,breakable}
\definecolor{main}{HTML}{D0D3D4}    
\definecolor{sub}{HTML}{D0D3D4}     
\newtcolorbox{dbox}{
    left=2pt,right=2pt,top=2pt,bottom=2pt,
    enhanced, 
    boxrule = 0pt,
    enlarge top by=5pt,
    enlarge bottom by=3pt,
  }
\tcbset{
    sharp corners,
    before skip = 0.1cm,    
    after skip = 0.01cm      
}

\def\summarybox#1#2{
\medskip
\begin{tcolorbox}[
  enhanced,
  title=#1,
  colframe=darkgray,
]
    #2
\end{tcolorbox}
}

\usepackage{tcolorbox}
\tcbuselibrary{skins,breakable}
\newtcolorbox{reviewcomment}[1][Comment]{  
    enhanced,
    frame hidden,
    borderline west={4pt}{0pt}{gray!50},
    colback=gray!10,
    boxrule=0pt,
    sharp corners,
    breakable,
    before upper={\parindent0pt\textbf{#1:}\par\vspace{0.3em}},
    fontupper=\itshape,
    left=10pt,
    right=10pt,
    top=5pt,
    bottom=5pt,
    before skip=1.5\baselineskip
}

\AtBeginDocument{%
  \providecommand\BibTeX{{%
    \normalfont B\kern-0.5em{\scshape i\kern-0.25em b}\kern-0.8em\TeX}}}

\setcopyright{cc}
\setcctype{by}
\acmJournal{TOSEM}
\acmYear{2026} \acmVolume{1} \acmNumber{1} \acmArticle{}
\acmMonth{1} \acmDOI{10.1145/3859242}

\begin{document}

\title{Agentic Refactoring: An Empirical Study of AI Coding Agents}

\author{Kosei Horikawa}
\email{horikawa.kosei.hk1@naist.ac.jp}
\orcid{0009-0006-6317-0754}
\affiliation{%
  \institution{Nara Institute of Science and Technology}
  \city{Ikoma}
  \country{Japan}
}

\author{Hao Li}
\affiliation{%
  \institution{Queen's University}
  \city{Kingston}
  \country{Canada}
}
\email{hao.li@queensu.ca}
\orcid{0000-0003-4468-5972}

\author{Yutaro Kashiwa}
\affiliation{%
  \institution{Nara Institute of Science and Technology}
  \city{Ikoma}
  \country{Japan}
}
\email{yutaro.kashiwa@is.naist.jp}
\orcid{0000-0002-9633-7577}

\author{Bram Adams}
\affiliation{%
  \institution{Queen's University}
  \city{Kingston}
  \country{Canada}
}
\email{bram.adams@queensu.ca}
\orcid{0000-0001-7213-4006}

\author{Hajimu Iida}
\affiliation{%
  \institution{Nara Institute of Science and Technology}
  \city{Ikoma}
  \country{Japan}
}
\email{iida@itc.naist.jp}
\orcid{0000-0002-2919-6620}

\author{Ahmed E. Hassan}
\affiliation{%
  \institution{Queen's University}
  \city{Kingston}
  \country{Canada}
}
\email{ahmed@cs.queensu.ca}
\orcid{0000-0001-7749-5513}

\renewcommand{\shortauthors}{Horikawa, et al.}

\begin{abstract}
Agentic coding tools, such as OpenAI Codex, Claude Code, and Cursor, are changing how software is developed. These AI-powered systems function as autonomous teammates capable of planning and executing complex development tasks. Agents have become active participants in refactoring, a key practice for improving internal code quality without altering observable behavior. Despite their increasing adoption, there is little empirical understanding of how agentic refactoring is used in practice, how it compares to human-driven refactoring, and what impact it has on code quality.

To address this empirical gap, we present a large-scale study of AI agent-generated refactorings in real-world open-source Java projects, analyzing \numint{\AgenticRefInstances} refactoring instances across \numint{\TotalPRs} pull requests and \numint{\TotalAgenticCommits} commits based on the AIDev dataset. Our empirical analysis shows that refactoring is a common and intentional activity in this development paradigm, with agents explicitly targeting refactoring in \AgentRefCommitPct\% of commits. The subsequent analyses of refactoring types, purposes, and code quality focus on the \numint{\AgenticRefCommits} commits (\numint{\SARInstances} instances) with explicit refactoring intent.
Analysis of refactoring types reveals that high-level refactorings form the largest category overall. 
However, agentic refactoring contains a slightly larger share of low-level, consistency-oriented edits than human refactoring (\AgenticLowPct{} vs. \HumanLowPct{}), and its two most frequent operations are \refactoring{Rename Variable} (\AgentRenameVariablePct{}) and \refactoring{Change Variable Type} (\AgentChangeVariableTypePct{}).
 Additionally, the stated motivations behind agentic refactoring focus overwhelmingly on internal quality concerns (with \ARMaintainabilityPct{} in maintainability and \ARReadabilityPct{} in readability), although the measured metric improvements from these frequently occurring low-level refactorings remain limited. 
Furthermore, quantitative evaluation of code quality metrics shows that agentic refactoring yields small but statistically significant improvements in structural metrics, particularly for medium-level changes, reducing class size and complexity.

\end{abstract}





\maketitle


\section{Introduction}\label{sec:introduction}
Refactoring is a cornerstone of sustainable software development that improves a software system's internal quality without changing its observable behavior~\cite{10.5555/169783}. Since Martin Fowler's catalog~\cite{DBLP:books/daglib/0019908} established the field's conceptual foundation, subsequent studies have emphasized refactoring as a core maintenance activity that supports long-term software evolvability. Prior work links refactoring to improved readability~\cite{DBLP:journals/tse/KimZN14,DBLP:conf/wcre/SellittoICLLPF22}, higher developer productivity~\cite{DBLP:conf/ifip2/MoserAPSS07}, and the removal of code smells~\cite{DBLP:conf/sigsoft/SilvaTV16}, as well as to preparing a codebase for future modifications~\cite{DBLP:journals/tosem/PantiuchinaZSPO20}. 

Despite its importance, refactoring demands deep domain knowledge and specialized skills. When executed improperly, refactoring can introduce bugs~\cite{DBLP:journals/tosem/PantiuchinaLZPL22}, break existing tests~\cite{DBLP:conf/icsm/RachatasumritK12, DBLP:conf/icsm/KashiwaS0BLKU21}, or destabilize the system~\cite{Bavota12refactorBug}. To mitigate these inherent risks, researchers have proposed many sophisticated approaches over the decades. These techniques include methods to identify appropriate refactoring locations \cite{DBLP:journals/tse/TsantalisC09}, refactoring recommendation systems \cite{DBLP:conf/kbse/MkaouerKBDC14}, and automatic refactoring approaches \cite{DBLP:journals/jss/ShahidiAN22}. While several studies reported that practitioners did not use refactoring tools~\cite{DBLP:journals/tse/Murphy-HillPB12}, the advent of large language models (LLMs) has recently accelerated these techniques and spurred practical adoption of automatic refactoring~\cite{DBLP:conf/apsec/ShirafujiOSMW23,DBLP:journals/eswa/DepalmaMHMA24,FraolICSME25}. Many studies report that practitioners now actively leverage LLMs to generate and apply refactoring operations~\cite{DBLP:conf/msr/TufanoMPDPB24,DBLP:conf/ease/WatanabeK0HYI24,DBLP:conf/msr/AlOmarVMNO24,DBLP:conf/msr/DeoHCWM24}.

Software engineering is now undergoing another shift with the rise of agentic coding tools~\cite{hassan2025agenticsoftwareengineering}. Unlike traditional prompt-based LLM workflows, where developers manually guide the AI step-by-step, agentic coding tools such as \codex,\footnote{\url{https://chatgpt.com/features/codex}} \claude,\footnote{\url{https://www.claude.com/product/claude-code}} and \cursor\footnote{\url{https://cursor.com/}} operate as AI teammates~\cite{DBLP:journals/corr/abs-2507-15003}. These agents can autonomously plan, execute, test, and iterate on complex development tasks with minimal human intervention~\cite{DBLP:journals/corr/abs-2505-19443}. In this new paradigm, agents act as collaborative refactoring teammates, actively participating in the refactoring process, ranging from simple cleanups to substantial design modifications. However, while these tools are being rapidly adopted, there is currently no empirical understanding of how this new class of agentic refactoring is used in practice, how it compares to human-driven refactoring, and what impact it has on code quality.

To address these critical gaps, we conduct a large-scale empirical study of refactorings produced in agentic coding workflows in real-world open-source projects based on the AIDev dataset~\cite{DBLP:journals/corr/abs-2507-15003}. We identify \numint{\AgenticRefInstances} refactoring instances across \numint{\TotalPRs} pull requests and \numint{\TotalCommits} commits in Java generated by agentic coding tools. We examine these agentic refactorings along four core dimensions (i.e., prevalence, types, purposes, and impacts) to answer the following research questions~(RQs):

\begin{itemize}[leftmargin=1em]
    \item \textbf{\rqA}
    To determine whether refactoring is a significant and intentional activity in agentic software development, we first quantify how often it occurs. We find that refactoring is common, with agents explicitly targeting it in \AgentRefCommitPct\% of agentic Java commits, which contain a total of \numint{\SARInstances} refactoring instances.
    
    \item \textbf{\rqB} 
    To assess the current sophistication of AI agents, it is crucial to determine whether they perform simple cosmetic cleanups or engage in more complex structural transformations. Our analysis reveals that, although high-level refactorings remain the largest abstraction-level category (\AgenticHighPct{}), agentic refactoring has a slightly larger share of low-level, consistency-oriented edits than human refactoring (\AgenticLowPct{} vs. \HumanLowPct{}).
    
    \item \textbf{\rqC} 
    Understanding the why behind these operations is essential to see if developers use agents for proactive quality improvement or for other tactical reasons. We find that agentic refactoring is overwhelmingly driven by the desire to improve internal code quality, specifically maintainability (\ARMaintainabilityPct) and readability (\ARReadabilityPct).
    
    \item \textbf{\rqD} 
    To justify agentic refactoring adoption, it is essential to quantitatively measure whether agentic refactoring provides tangible benefits to the codebase's structural health. The results show that agentic refactoring yields small but significant improvements in structural metrics like code size and complexity, yet it currently fails to consistently reduce the overall count of known design and implementation smells.
\end{itemize}

Our findings provide the first large-scale empirical baseline of agentic refactoring. These results have direct implications for developers, helping them set expectations for what current agents can (and cannot) reliably refactor. For researchers, our work enables new studies of human-agent collaboration in software maintenance. Finally, for coding agent builders, our findings highlight current limitations
 (\eg limited impact on design smells) and opportunities for creating agents that can perform more sophisticated, structurally-aware refactorings.

\smallskip
\textbf{Replication Package.} To facilitate replication and further studies, we provide the data used in our replication package.\footnote{\url{https://github.com/Mont9165/Agent_Refactoring_Analysis}}

\smallskip
\textbf{Paper Structure.} The remainder of this paper is organized as follows. Section~\ref{sec:relatedwork} reviews related work. Section~\ref{sec:motivation} presents the motivating examples. Section~\ref{sec:studydesign} describes how we collected and filtered the data. Section~\ref{sec:results} presents the findings of our research questions. Section~\ref{sec:implications} summarizes the key insights and discusses the implications of our study. Section~\ref{sec:threats_to_validity} outlines threats to validity, and Section~\ref{sec:conclusion} concludes the paper.

\section{Related Work}
\label{sec:relatedwork}

In this section, we discuss related work about the foundations of refactoring, large-scale empirical analyses, automated refactoring techniques, and AI-assisted software development.

\subsection{Refactoring and Its Empirical Foundations}
Research on refactoring has built a substantial empirical foundation since Fowler cataloged its patterns~\cite{DBLP:books/daglib/0019908}. Beyond establishing refactoring as a core maintenance practice~\cite{DBLP:journals/eInformatica/WilkingKK07, DBLP:conf/ifip2/MoserAPSS07, DBLP:conf/icse/KimCK11, DBLP:conf/sbes/ChavezFFCG17, DBLP:journals/tse/DallalA18}, this line of work has examined why developers refactor: to improve readability, reduce duplication, and facilitate the maintainability of software systems~\cite{DBLP:conf/sigsoft/SilvaTV16, DBLP:conf/sigsoft/KimZN12}, but also to improve comprehension, reduce cognitive load, and prepare code for future changes~\cite{DBLP:journals/tse/DallalA18}. For instance, Kim\etal~\cite{DBLP:journals/tse/KimZN14} surveyed Microsoft developers and found that refactoring is widely regarded as beneficial for improving code quality and productivity. Specifically, developers cited the benefits of refactoring as improved readability, improved maintainability, and fewer bugs. Their quantitative analysis of Windows 7 modules also showed that a dedicated refactoring effort led to measurable benefits, including a significant reduction in inter-module dependencies and post-release defects. Their findings provide the conceptual basis for our analysis of refactoring purpose in agentic commits (RQ3).
Building on these foundational insights, subsequent research has investigated refactoring practices at scale to understand their prevalence and impact on software quality.

\subsection{Large-Scale Refactoring Studies and Quality Impact}
The large-scale mining of refactorings has revealed that refactoring is pervasive in both industrial and open-source development~\cite{Zhenchang2006, DBLP:journals/tse/Murphy-HillPB12, DBLP:journals/scp/VassalloGPGB19}. 
More recently, large-scale refactoring detectors such as \textsc{RefactoringMiner}~\cite{DBLP:journals/tse/TsantalisKD22, DBLP:journals/tosem/AlikhanifardT25} and \textsc{RefDiff}~\cite{DBLP:journals/tse/SilvaSSTV21} have enabled mining-based studies at scale. 

Although refactoring is generally expected to improve quality, empirical evidence on its actual impact remains mixed. 
Some studies report positive effects on code stability, readability, and productivity, particularly in agile settings~\cite{icsme/horikawa25, DBLP:conf/sigsoft/KimZN12, DBLP:conf/ifip2/MoserAPSS07}, 
whereas others show that refactoring does not consistently remove design problems or prevent defects~\cite{DBLP:journals/jss/BavotaLPOP15}. 
For example, Cedrim\etal~\cite{DBLP:conf/sigsoft/CedrimGMGSMFRC17} found that less than 10\% of refactorings effectively remove code smells, while over 30\% introduce new ones. 
Similarly, Szoke\etal~\cite{DBLP:conf/scam/SzokeANFG14} observed that only large-scale, systematic bulk refactorings lead to measurable quality improvement, suggesting that scope matters as much as type. Bavota\etal~\cite{Bavota12refactorBug} further noted that inheritance-related refactorings are especially error-prone, emphasizing the need to consider technical context.

Beyond the technical dimension, developer intent matters. 
Palomba\etal~\cite{DBLP:conf/iwpc/PalombaZOL17} showed that refactorings frequently co-occur with bug fixes, implying opportunistic rather than proactive behavior. 
This aligns with Silva\etal~\cite{DBLP:conf/sigsoft/SilvaTV16} and Pantiuchina\etal~\cite{DBLP:journals/tosem/PantiuchinaZSPO20}, who found that evolving requirements, readability, and maintainability, rather than smell removal, drive most refactoring decisions. 
Taken together, these findings highlight that refactoring impact depends on its type, scope, and motivation. We extend this perspective to agentic refactoring in RQ2 and RQ4.

\subsection{Automated Refactoring Support}
Automated and semi-automated refactoring has been explored in software engineering for over two decades. Rule-based tools such as \textsc{JDeodorant}~\cite{DBLP:conf/wcre/TsantalisCC18,DBLP:conf/icse/MazinanianTSV16} and search-based refactoring engines~\cite{DBLP:journals/jss/OKeeffeC08} attempted to reduce manual effort; however, they suffered from low adoption due to lack of trust and limited semantic reasoning~\cite{DBLP:journals/tse/Murphy-HillPB12,DBLP:journals/scp/VassalloGPGB19}. 
Recently, the refactoring capability of large language models (LLMs) has also been explored~\cite{DBLP:journals/eswa/DepalmaMHMA24, DBLP:journals/corr/abs-2411-02320}. Cordeiro\etal~\cite{DBLP:journals/corr/abs-2411-02320} evaluated the refactoring quality of StarCoder2 under various prompt-engineering strategies (\eg zero-shot, one-shot) and found that LLMs can outperform developers in removing code smells under specific prompting configurations. However, these studies focus on \textit{prompt-based refactoring}, where models perform one-shot transformations given explicit instructions.

In contrast, our study explores \textit{agentic refactoring}, where autonomous coding agents (\eg~\codex, \devin, \cursor) plan, execute, and validate changes through iterative reasoning and feedback. This paradigm differs from prompt-based generation in that agents can decompose complex objectives, perform refactoring alongside other maintenance activities, and generate verifiable pull requests with minimal human intervention. Thus, our work provides the first large-scale empirical view of intentional, agentic refactoring practices carried out through agentic coding workflows.


\subsection{AI-Assisted Software Development}
AI-based coding assistance has evolved rapidly from autocomplete-style tools~\cite{DBLP:conf/sigsoft/SvyatkovskiyDFS20} to modern LLM-based copilots capable of nontrivial synthesis and transformation~\cite{DBLP:journals/jss/DakhelMNKDJ23}. Empirical studies report that AI-generated code accelerates development but often introduces maintainability, redundancy, and security concerns~\cite{DBLP:conf/uss/SandovalPNKGD23,DBLP:journals/ese/AsareNA23}. 
Trust and validation remain key challenges when adopting AI-generated code. 
Sergeyuk\etal~\cite{DBLP:journals/infsof/SergeyukGBA25} reported that developers frequently verify AI suggestions manually due to concerns over correctness and control. 
In particular, 21.9\% of respondents avoided using AI for refactoring tasks, highlighting that trust is critical where functional behavior must remain unchanged. LLMs further exhibit inconsistency and limited contextual understanding~\cite{DBLP:conf/uss/SandovalPNKGD23, DBLP:journals/ese/AsareNA23}. Even with identical prompts, tools like ChatGPT may produce divergent or unnecessary edits, increasing review effort and cognitive overhead~\cite{DBLP:journals/eswa/DepalmaMHMA24}.
Evidence from agentic coding platforms further supports these findings. 
A recent study of autonomously generated pull requests (Agentic-PRs)~\cite{DBLP:journals/corr/abs-2509-14745} found that 45.1\% required post-review fixes (most often for bugs, refactoring, or documentation), despite agents performing refactoring more frequently than humans (24.9\% vs.~14.9\%). 
Overall, while AI-assisted tools can restructure code effectively, human oversight remains essential to ensure correctness and maintainability. This issue becomes even more significant in \textit{agentic refactoring}, where AI autonomously plans and applies code changes.


\subsection{Agentic Software Development and Research Gap}
Recent work has introduced the notion of \textit{agentic software engineering}~\cite{hassan2025agenticsoftwareengineering}, where AI systems act as autonomous collaborators capable of proposing, modifying, and integrating pull requests~\cite{DBLP:journals/corr/abs-2507-15003}. The AIDev dataset enables empirical research on such activities at scale. However, while early work analyzes AI contributions in issue resolution~\cite{DBLP:conf/msr/ChouchenBBOAM24} and documentation~\cite{DBLP:conf/icsm/RahmanZMRRS24}, and reports refactoring only as one broad category of agent activity, no study has mined the concrete refactoring operations that agents perform or examined how AI-generated refactorings differ in type, intent, and impact from human refactoring. Our work addresses this gap by providing the first large-scale, operation-level empirical analysis of refactoring in agentic commits, introducing \textit{Agentic Refactoring} as a novel concept and dataset.

In summary, prior research has matured along multiple refactoring dimensions (motivation, automation, empirical evolution, and human factors), but agentic refactoring remains unexplored. Our study is the first to provide a large-scale empirical view of agentic refactoring by (i) quantifying its operation-level prevalence, (ii) contrasting AI and human refactoring styles, (iii) classifying refactoring intent in AI settings, and (iv) evaluating its quality impact.

\section{Motivating Example}\label{sec:motivation}

AI agents are increasingly integrated into software development workflows. Coding agents such as Claude Code, Devin, and Cursor can not only generate code but also collaboratively refactor existing codebases with developers~\cite{DBLP:journals/corr/abs-2509-14745}. Rather than acting as fully autonomous systems, these agents increasingly function as collaborative refactoring teammates, assisting humans in restructuring large or complex code to enhance maintainability and readability.

\begin{figure}[t]
\centering
  \begin{subfigure}[t]{0.95\linewidth}
    \centering
    \includegraphics[width=\linewidth]{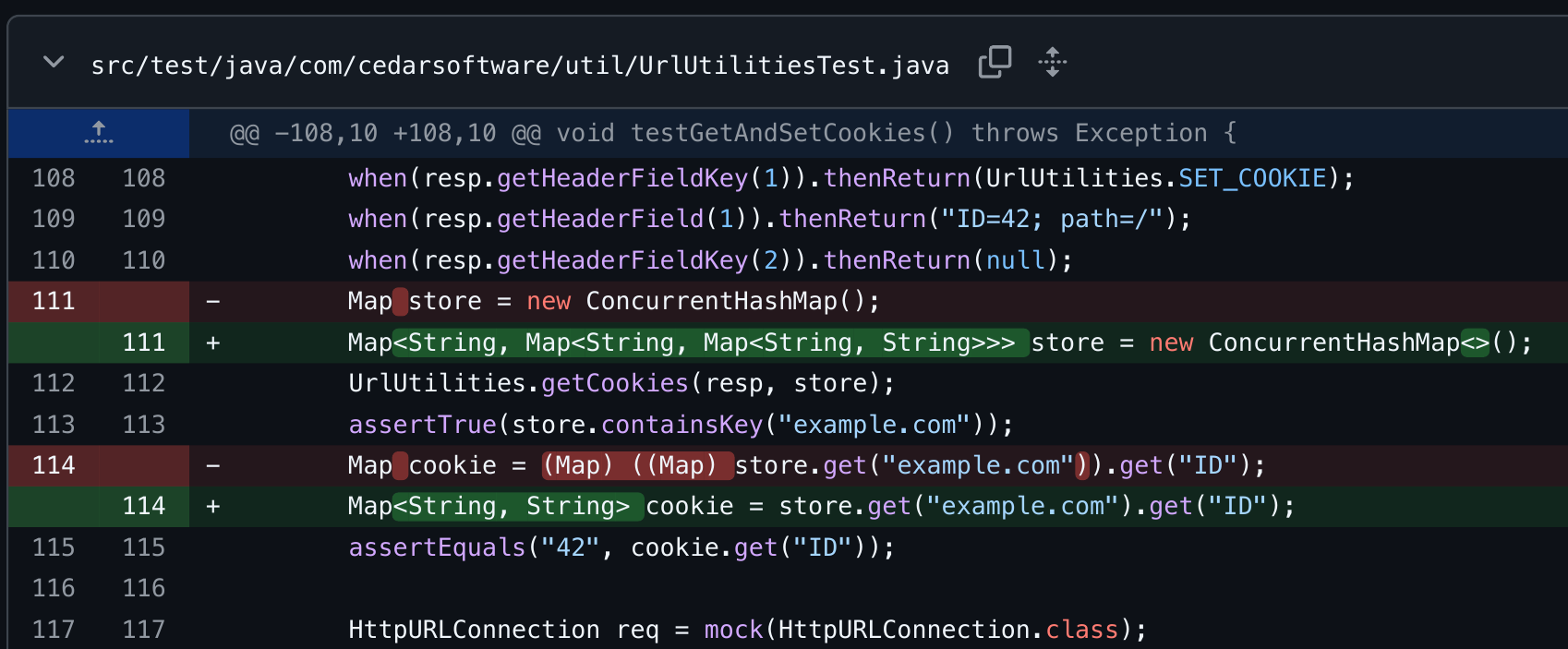}
    \caption{ Replacing raw types with parameterized types to improve type safety. }
    \label{fig:refactor_change_variable}
  \end{subfigure}
  \vspace{0.5cm}
  \vfill
  \begin{subfigure}[t]{0.95\linewidth}
    \centering
    \includegraphics[width=\linewidth]{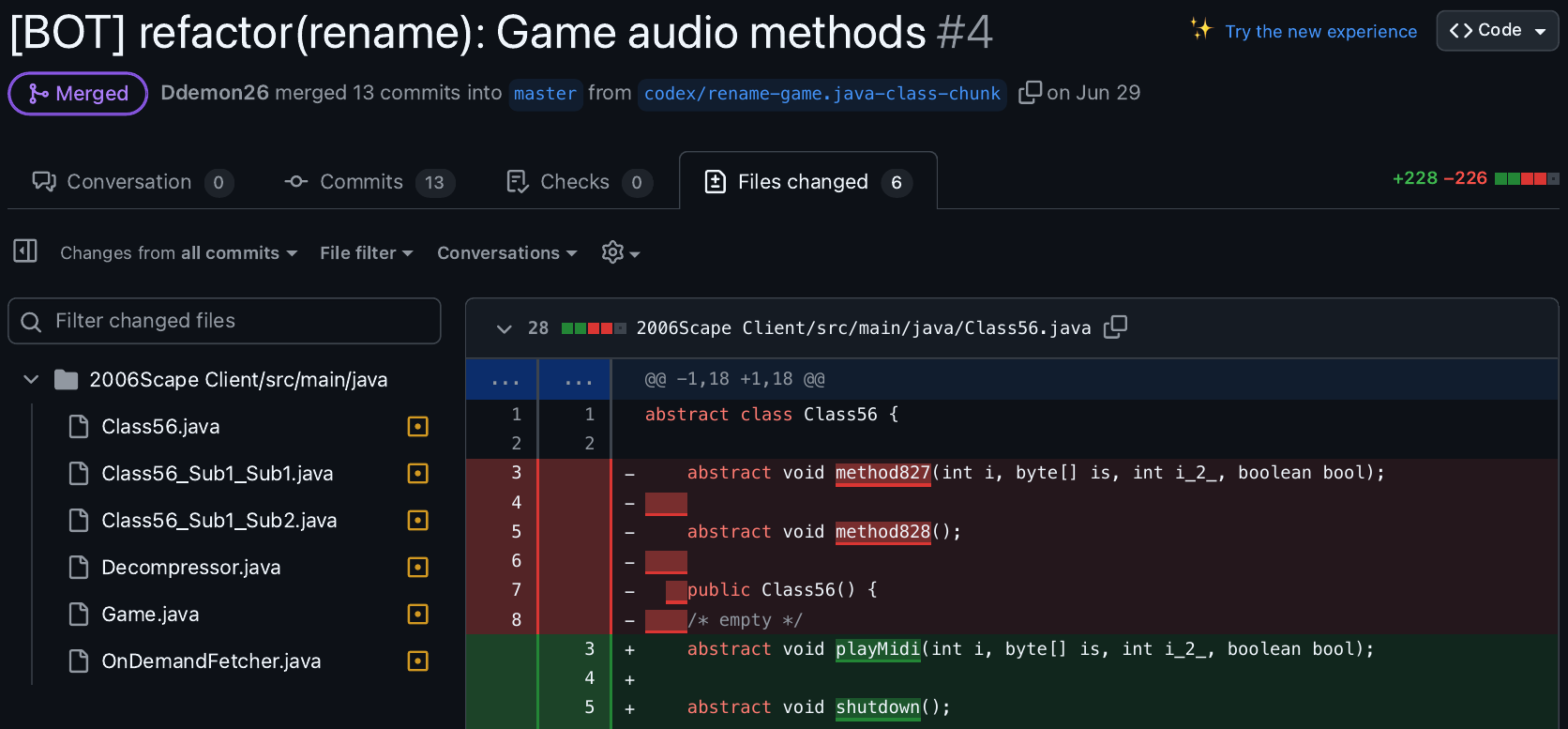}
    \caption{Standardizing method names across multiple files.}
    \label{fig:bot_refactor_rename}
  \end{subfigure}
\caption{Examples of agentic refactoring.}
\label{fig:agentic_refactor_examples}
\end{figure}

Figure~\ref{fig:agentic_refactor_examples} illustrates two representative examples of such agentic refactoring. Figure~\ref{fig:refactor_change_variable} demonstrates an agent automatically replacing raw types with parameterized generic types to improve type safety.\footnote{\url{https://github.com/jdereg/java-util/commit/15b8e6d932ead9860f854bd0e0e84e16f2bbf230}} By systematically converting raw \texttt{Map} declarations to their fully parameterized equivalents, the agent eliminates unsafe casts and improves code clarity. 
In contrast, Figure~\ref{fig:bot_refactor_rename} shows another type of agentic refactoring, where an agent systematically renames methods to improve consistency.\footnote{\url{https://github.com/Ddemon26/2006Scape/pull/4/files}} Such operations improve naming clarity and stylistic uniformity but have little direct effect on structural quality metrics.

These contrasting examples show that AI agents are active participants in the refactoring process. However, this observation raises a critical question: Do these operations consistently improve software quality metrics (\eg lines of code, cyclomatic complexity, coupling), or do they merely restructure code superficially? Despite the rapid adoption of these agents, there remains limited empirical understanding of how such agentic refactorings are used in practice.

To address this gap, we conduct a large-scale empirical study of agentic refactorings in real-world open-source projects. We systematically examine four dimensions of agentic refactoring:  
the \textbf{frequency} of AI participation (RQ1),  
the \textbf{types} of refactoring operations performed (RQ2),  
the \textbf{purpose} expressed by developers and AI agents (RQ3),  
and the resulting \textbf{impacts} on internal quality metrics (RQ4).  
Together, these research questions aim to clarify whether agentic refactorings merely automate surface-level cleanup tasks or contribute to structural and maintainability improvements in software systems.

\begin{figure}[t]
\centering
\includegraphics[width=\linewidth]{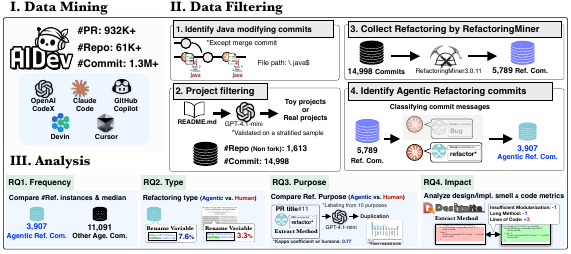}
\caption{Overview of the study design.}
\label{fig:overview}
\end{figure}

\section{Data Collection}\label{sec:studydesign}

In this section, we describe the data mining and data filtering as outlined in Figure~\ref{fig:overview}.

\subsection{Data Mining}\label{subsec:data_mining}

To study agentic refactoring in real-world projects, we start with the AIDev dataset~\cite{DBLP:journals/corr/abs-2507-15003}. This dataset is an ideal source as it contains 932,791 pull requests (PRs) created by five coding agents from over 61,000 repositories. 
However, the commit data included in the AIDev dataset\footnote{\url{https://huggingface.co/datasets/hao-li/AIDev}} (the \texttt{pr\_commits} table) only covers repositories with more than 100 GitHub stars.
To ensure broader coverage, we leverage the GitHub REST API\footnote{\url{https://docs.github.com/en/rest}} to mine all commits using the \texttt{all\_pull\_request} table, which provides pull request metadata for all repositories regardless of star count. In total, we collect 1,311,057 agentic commits and share this expanded dataset publicly as AIDev-full.\footnote{\url{https://huggingface.co/datasets/hao-li/AIDev-full}}

\subsection{Data Filtering}

To construct a focused dataset tailored for refactoring analysis, we perform a multi-stage filtering process based on the collected data curated from the AIDev dataset.

\subsubsection{Identify Java Modifying Commits}

 In this study, we focus on commits that modify at least one \texttt{.java} file for three reasons. 
First, focusing on Java is consistent with the majority of large-scale empirical refactoring studies~\cite{DBLP:journals/tse/Murphy-HillPB12,DBLP:journals/scp/VassalloGPGB19,DBLP:conf/sigsoft/CedrimGMGSMFRC17}, enabling direct comparison with prior work. 
Second, \RMiner~\cite{DBLP:journals/tse/TsantalisKD22, DBLP:journals/tosem/AlikhanifardT25} is the only refactoring detection tool that supports 103 types with an F-score of 99.5\%; multi-language alternatives such as RefDiff~\cite{DBLP:journals/tse/SilvaSSTV21} support substantially fewer types. 
Third, \designitejava~\cite{DBLP:conf/msr/Sharma24} provides comprehensive code quality metrics, code smell detection, and design smell detection specifically for Java, enabling the multi-perspective analysis in RQ4.

We further exclude all merge commits to avoid duplicated changes and ambiguous parent relationships. 

\subsubsection{Project Filtering}
To ensure that our analysis targets substantive software systems rather than educational or trivial repositories, we apply several steps to filter the projects.

\paragraph{Step 1 -- Automated Classification.}

We leverage GPT-4.1-mini to classify each repository based on the content of its README.md file. We selected GPT-4.1-mini for its cost efficiency given the scale of the dataset (\numint{\ReposBeforeFiltering} repositories); the total API cost was approximately \GPTCostTotal{} for \GPTCostTokens{} tokens, whereas using a higher-end model (\eg Claude Opus) would cost approximately \$120 for the same task. The classification prompt was designed to operationalize the three category definitions described below and was verified on examples before application to the full dataset. The full prompt and its development process are documented in the replication package.\footnote{\url{https://github.com/Mont9165/Agent_Refactoring_Analysis/blob/main/PROMPTS.md}}  
The classification reliability is ensured through human validation (see Step~2). 
Each project is assigned to one of the following three categories:

\begin{itemize}

    \item \texttt{non\_toy}: Software repositories with meaningful functionality intended for real-world use or research purposes (\eg web frameworks, database client libraries, build plugins, academic research tools, domain-specific utilities), identified by the presence of usage instructions, installation guides, CI pipelines, versioned releases, documented APIs, or active issue management.

    \item \texttt{toy}: Tutorial repositories, coursework material, evaluation harnesses, or minimal demonstrative examples, identified by explicit indicators such as assignment instructions, tutorial structure, absence of CI/CD, no versioned releases, or no evidence of external usage.
    
    \item \texttt{uncertain}: Repositories whose purpose cannot be reliably determined due to minimal or missing README.md, ambiguous descriptions, or insufficient information to classify as either \texttt{non\_toy} or \texttt{toy}.

\end{itemize}

\paragraph{Step 2 -- Manual Verification.}
To validate the reliability of the automated classification in Step~1, we conducted a systematic manual verification. We drew a stratified proportional sample of \ValidationSampleTotal{} repositories from the full set of \numint{\ReposBeforeFiltering} GPT-classified repositories (95\% confidence level, 5\% margin of error, using the Cochran formula). Of the \ValidationSampleTotal{} sampled repositories, \ValidationSampleExcluded{} were excluded because the repositories had been deleted or made private since the time of data collection, leaving \ValidationSampleAnalyzed{} for analysis. Two annotators (with \ValidationAnnotatorOneExp{} and \ValidationAnnotatorTwoExp{} years of programming experience), blind to the GPT-4.1-mini labels, independently classified each of the \ValidationSampleAnalyzed{} repositories based on its README, source code, and GitHub activity. Disagreements were resolved through discussion with a third annotator (\ValidationAnnotatorThreeExp{} years of experience). As shown in Table~\ref{tab:project_filtering_validation}, inter-rater agreement between annotators was Cohen's $\kappa$ = \ValidationHumanKappa{} (almost perfect agreement), and agreement between GPT-4.1-mini and the adjudicated human labels was Cohen's $\kappa$ = \ValidationGPTKappa{} (substantial agreement). These results confirm that the automated classification is consistent with human judgment. 

To characterize the disagreements behind these kappa scores, we inspected all 32 of the 319 cases in which the GPT-4.1-mini label differed from the adjudicated human label. All 32 disagreements involved the boundary between \texttt{non\_toy} and \texttt{toy}, and most of them followed the same direction. In 24 of the 32 cases, GPT-4.1-mini labeled a repository as \texttt{non\_toy}, while the annotators labeled it as \texttt{toy}. A typical case was a personal or experimental project with a README that described substantive functionality, such as a microservice-based terminal-operations system or an AI question-answering service. However, the repository had no releases, no CI, and no evidence of external use. The annotators also inspected the source code and GitHub activity. They gave these signals more weight than the README.
The reverse error occurred in only 8 of the 32 cases. These cases usually involved niche but functional software, such as a game-specific mod with installation and troubleshooting instructions. The same boundary accounts for all 26 disagreements between the two annotators. This result suggests that distinguishing \texttt{non\_toy} from \texttt{toy} is the most difficult judgment in this task.
For our study, this error pattern has a limited effect. About 7.5\% of the repositories (24 of 319 cases) may be toy-like projects that pass the filter and add noise. By contrast, substantive projects are rarely excluded.

\begin{table}[t]
\centering
\caption{Reliability of project filtering classification}
\label{tab:project_filtering_validation}
\begin{tabular}{lrrr}
\toprule
Comparison & Cohen's $\kappa$ & Accuracy & Macro F1 \\
\midrule
Human vs Human        & \ValidationHumanKappa & \ValidationHumanAgreement & \ValidationHumanMacroF \\
GPT-4.1-mini vs Human & \ValidationGPTKappa   & \ValidationGPTAgreement   & \ValidationGPTMacroF \\
\bottomrule
\end{tabular}
\end{table}

\paragraph{Step 3 -- Final Project Selection.}
 After classification and verification, we retain \numint{\NonToyProjectNum} projects labeled as \texttt{non\_toy}. We exclude \numint{\toyProjectNum} repositories categorized as \texttt{toy} and \numint{\uncertainProjectNum} as \texttt{uncertain}. 
Finally, to avoid redundancy, we remove forked repositories. This results in a final corpus of \numint{\TotalRepos} unique software projects for analysis. The Java subset includes \numint{\TotalCommits} non-merge commits suitable for automated refactoring detection and metric extraction.
This filtering step reduces noise from trivial repositories and ensures that our analysis focuses on substantive software systems.

\subsubsection{Collect Refactoring Operations by \RMiner}
To identify specific refactoring operations within our set of Java commits, we use \RMiner 3.0\footnote{\url{https://github.com/tsantalis/RefactoringMiner/releases/tag/3.0.11}} for each commit. \RMiner detects 103 distinct refactoring types and identifies their precise locations, achieving an overall F-score of 99.5\%~\cite{DBLP:journals/tse/TsantalisKD22, DBLP:journals/tosem/AlikhanifardT25}.
Specifically, \RMiner analyzes the abstract syntax trees (ASTs) of the modified Java files between consecutive revisions to identify the types and locations of applied refactorings. 

We adopt the commit as the unit of analysis in this study. \RMiner operates on pairs of consecutive revisions, making the commit the natural and necessary unit for refactoring detection. This is consistent with prior large-scale empirical refactoring studies~\cite{DBLP:journals/tse/Murphy-HillPB12,DBLP:journals/scp/VassalloGPGB19,DBLP:conf/sigsoft/CedrimGMGSMFRC17}, which also adopt the commit as their unit of analysis. Furthermore, the scale of our dataset (\numint{\TotalCommits} commits across \numint{\TotalRepos} repositories) requires automated and reproducible methods. 
This process generates a dataset linking specific commits to the refactoring operations they contain.
From the initial set of \numint{\TotalCommits} commits, \RMiner identifies \numint{\RefCommits} commits as containing at least one refactoring operation.

\begin{table}[t]
    \centering
    \caption{List of Self-Affirmed Refactoring (SAR) Patterns from AlOmar\etal~\cite{DBLP:conf/icse/AlOmarM019}}
    \label{tab:sar_patterns}
    \scriptsize
    \begin{tabularx}{\linewidth}{lX lX lX}
        \toprule
        (1) Refactor* & & (30) Removed poor coding practice & & (59) Change design & \\
        (2) Mov* & & (31) Improve naming consistency & & (60) Modularize the code & \\
        (3) Split* & & (32) Removing unused classes & & (61) Code cosmetics & \\
        (4) Fix* & & (33) Pull some code up & & (62) Moved more code out of & \\
        (5) Introduce* & & (34) Use better name & & (63) Remove dependency & \\
        (6) Decompos* & & (35) Replace it with & & (64) Enhanced code beauty & \\
        (7) Reorganiz* & & (36) Make maintenance easier & & (65) Simplify internal design & \\
        (8) Extract* & & (37) Code cleanup & & (66) Change package structure & \\
        (9) Merg* & & (38) Minor Simplification & & (67) Use a safer method & \\
        (10) Renam* & & (39) Reorganize project structures & & (68) Code improvements & \\
        (11) Chang* & & (40) Code maintenance for refactoring & & (69) Minor enhancement & \\
        (12) Restructur* & & (41) Remove redundant code & & (70) Get rid of unused code & \\
        (13) Reformat* & & (42) Moved and gave clearer names to & & (71) Fixing naming convention & \\
        (14) Extend* & & (43) Refactor bad designed code & & (72) Fix module structure & \\
        (15) Remov* & & (44) Getting code out of & & (73) Code optimization & \\
        (16) Replac* & & (45) Deleting a lot of old stuff & & (74) Fix a design flaw & \\
        (17) Rewrit* & & (46) Code revision & & (75) Nonfunctional code cleanup & \\
        (18) Simplifi* & & (47) Fix technical debt & & (76) Improve code quality & \\
        (19) Creat* & & (48) Fix quality issue & & (77) Fix code smell & \\
        (20) Improv* & & (49) Antipattern bad for performances & & (78) Use less code & \\
        (21) Add* & & (50) Major/Minor structural changes & & (79) Avoid future confusion & \\
        (22) Modif* & & (51) Clean up unnecessary code & & (80) More easily extended & \\
        (23) Enhanc* & & (52) Code reformatting \& reordering & & (81) Polishing code & \\
        (24) Rework* & & (53) Nicer code / formatted / structure & & (82) Move unused file away & \\
        (25) Inlin* & & (54) Simplify code redundancies & & (83) Many cosmetic changes & \\
        (26) Redesign* & & (55) Added more checks for quality factors & & (84) Inlined unnecessary classes & \\
        (27) Cleanup & & (56) Naming improvements & & (85) Code cleansing & \\
        (28) Reduc* & & (57) Renamed for consistency & & (86) Fix quality flaws & \\
        (29) Encapsulat* & & (58) Refactoring towards nicer name analysis & & (87) Simplify the code & \\
        \bottomrule
    \end{tabularx}
\end{table}

\subsubsection{Identify \AR Commits}
 To understand how often agentic refactoring is an intentional act, we must identify commits that explicitly state refactoring intent. We define a commit as having \emph{explicit refactoring intent} if (i)~its commit message matches at least one of the Self-Affirmed Refactoring (SAR) keyword patterns from AlOmar\etal~[2], and (ii)~\RMiner detects at least one refactoring operation in the commit. We refer to such commits as \emph{Agentic Refactoring} commits. The keyword patterns (listed in Table~\ref{tab:sar_patterns}) were developed and validated by AlOmar \etal and have been widely adopted in self-affirmed refactoring research; our study applies these established patterns without modification.

We distinguish \emph{intent} from \emph{purpose} in this study. Intent is a binary classification of whether a commit explicitly targets refactoring (this section, based on SAR keyword matching). Purpose is a multi-class classification of why the refactoring was performed (\eg maintainability, readability), analyzed in Section~5.3.2 using GPT-4.1-mini.

The complete list of patterns is presented in Table~\ref{tab:sar_patterns}. 
Applying this procedure to the \numint{\TotalCommits} commits yields \numint{\SARCommits} \texttt{Agentic Refactoring Commits}.

\begin{table}[t]
\centering
\caption{Distribution of commits and PRs by associated AI agent.}
\label{tab:agent-distribution}
\begin{tabular}{lrr}
\toprule
AI Agent  & \# PRs (\%)& \# Commits (\%) \\
\midrule
OpenAI Codex & \numint{\CodexPRs} (\CodexPRPct) & \numint{\CodexCommits} (\CodexCommitPct) \\
Devin & \numint{\DevinPRs} (\DevinPRPct) & \numint{\DevinCommits} (\DevinCommitPct) \\
Cursor  & \numint{\CursorPRs} (\CursorPRPct) & \numint{\CursorCommits} (\CursorCommitPct)\\

Claude Code  & \numint{\ClaudePRs} (\ClaudePRPct) & \numint{\ClaudeCommits} (\ClaudeCommitPct)\\
\midrule
Total & \numint{\TotalPRs} (100\%) & \numint{\TotalCommits} (100\%) \\
\bottomrule

\multicolumn{3}{l}{\scriptsize  *GitHub Copilot is part of the AIDev dataset but produced no commits after filtering, so it is excluded here. }
\end{tabular}
\end{table}

\subsubsection{Dataset Overview}\label{subsec:dataset_overview}
After filtering, our curated dataset contains \numint{\TotalCommits} unique, non-merge commits that modify at least one Java file. Of these, \numint{\SARCommits} are labeled as \texttt{agentic refactoring commits} (those with explicit refactoring intent in their messages) and the remaining \numint{\OthersCommits} commits are categorized as \texttt{Other Agentic Commits}.

These \numint{\TotalCommits} commits originate from \numint{\TotalPRs} PRs across \numint{\TotalRepos} repositories. Among these PRs, \numint{\ClosedPRs} (\ClosedPRsPct) are closed and \numint{\MergedPRs} (86.9\%) are merged. This high merge rate indicates that most agentic contributions were integrated into their respective projects, demonstrating substantial acceptance of agent-generated code. Regarding agent participation, the distribution across the agents is shown in Table~\ref{tab:agent-distribution}. Although the AIDev dataset includes five coding agents, GitHub Copilot produced no Java-modifying, non-merge commits after our filtering and is therefore absent from our final dataset. 
The data reveal that \codex dominates the dataset, accounting for \CodexCommitPct of all commits and \CodexPRPct of PRs. \devin\ and \cursor\ contribute \DevinCommitPct and \CursorCommitPct respectively, while \claude\ accounts for only \ClaudeCommitPct of the commits analyzed.

\section{Results}\label{sec:results}
In this section, we provide the motivation, approach, and findings for each of our research questions. Figure~\ref{fig:overview} provides an overview of our study design.

\subsection{\rqA}\label{sec:rqa}

\subsubsection{Motivation}
Agentic coding is transitioning from a novelty to an everyday practice, yet little is known about the extent of refactoring performed by these agents in real-world projects. Although agentic coding tools promise to act as autonomous collaborators rather than mere code generators~\cite{hassan2025agenticsoftwareengineering}, most research has focused on their ability to create new features or fix bugs. Their role in software maintenance, particularly refactoring, remains largely unexplored. Before assessing the quality or types of refactorings agents perform, we must first establish a baseline: Is agentic refactoring a significant real-world activity or merely an incidental byproduct of other development tasks? Quantifying its frequency is the first step toward understanding whether these agents contribute to refactoring.

\subsubsection{Approach}

We record the distribution of refactoring instances for the collected \numint{\TotalCommits} commits. Both groups in this comparison consist exclusively of agentic commits: \emph{Agentic Refactoring} denotes commits with both detected refactoring and explicit refactoring intent in the commit message, while \emph{Other Agentic Commits} denotes all remaining agentic commits. No human commits are included. To determine whether the distributions of refactoring instances differ between these two groups, we perform the Mann--Whitney U test~\cite{Mann1947OnAT}, a non-parametric rank-based test that does not assume normality, at a significance level of $\alpha=0.05$. 
We also compute Cliff’s delta~$d$~\cite{Cliff} effect size to quantify the difference based on the following thresholds~\cite{Cliff_threshold}:

\begin{equation} \label{effectsize}
\mathrm{Effect \ size} = 
\left\{
\begin{array}{ll}
	negligible,  & \mathrm{if} \ |d|  \le 0.147 \\
	small,  & \mathrm{if} \ 0.147 < |d|  \le 0.33 \\
	medium,  & \mathrm{if} \ 0.33 < |d|  \le 0.474 \\
	large,  & \mathrm{if} \ 0.474 < |d|  \le 1 \\
\end{array}\right.
\end{equation}

We examine refactoring activity from two complementary perspectives: (i) aggregate prevalence, based on the total number of detected refactoring instances in each group, and (ii) per-commit prevalence, based on the distribution of refactoring instances per commit.


\begin{table}[t]
\centering
\caption{Summary of detected refactoring instances in agentic commits.}
\label{tab:rq1-frequency}
\begin{tabular}{lrr}
\toprule
 & \# Commits (\%) & \# Instances (\%) \\ 
\midrule
Agentic Refactoring & \numint{\SARCommits} (\AgentRefCommitPct \%) & \numint{\SARInstances} (\SARInstancePct \%) \\ 
Other Agentic Commits  & \numint{\OthersCommits} (\OthersCommitsPct \%) & \numint{\NonSARInstances} (\NonSARInstancesPct \%) \\ 
\midrule
Total & \numint{\TotalCommits} (100\%)&  \numint{\AgenticRefInstances}(100\%)\\

\bottomrule
\end{tabular}
\end{table}






\subsubsection{Findings}
\finding{Refactoring is common in agentic software development and often appears even without explicit intent.}  
As shown in Table~\ref{tab:rq1-frequency},  commits with explicit refactoring intent account for \AgentRefCommitPct\% (\numint{\AgenticRefCommits} out of \numint{\TotalCommits}) of all commits and contain \textbf{\numint{\SARInstances}} detected refactoring instances. However, the remaining \numint{\OthersCommits} other agentic commits~(whose messages do not explicitly indicate refactoring) contain \numint{\NonSARInstances} detected refactoring instances (\NonSARInstancesPct\% of all instances). This pattern suggests tangled commits: while implementing a feature or fixing a bug, agents also refactor nearby code (\eg rename variables, extract small helpers) within the same commit.

\begin{figure}[tb]
    \centering
    \includegraphics[width=0.55\linewidth]{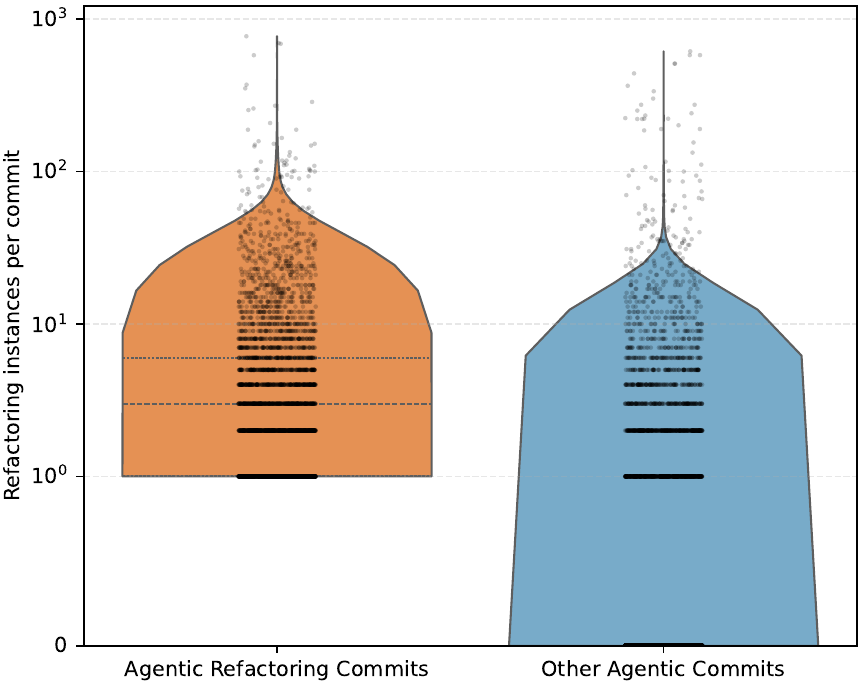}
    \caption{Distribution of refactoring instances per commit (agentic vs. other agentic).}
    \label{fig:rq1_ref_instances}
\end{figure}

\smallskip
\noindent\finding{When agents state refactoring intent, they perform significantly more refactoring instances than other agentic commits.}
Figure~\ref{fig:rq1_ref_instances} shows the distribution of refactoring instances per commit in agentic refactoring and  other agentic commits. Although other agentic commits contribute many refactoring instances in aggregate, those instances are spread across a much larger number of commits.  The result of the Mann--Whitney $U$ test shows that the difference is statistically significant ($p\le0.001$). In addition, the result of Cliff's delta ($d=0.838$) indicates a large effect size. The results suggest that when AI agents explicitly signal refactoring intent, they perform more concentrated and substantial restructuring activities. In contrast, when refactoring is unacknowledged, it tends to appear sporadically or incidentally as part of broader code modifications, such as feature implementation or bug fixing.

\summarybox{\textbf{Answer to RQ1}}{
Refactoring is a common and intentional activity in agentic software development, with \AgentRefCommitPct\% of agentic commits explicitly targeting refactoring.
}

\subsection{\rqB}\label{sec:rqb}

\subsubsection{Motivation}
Refactoring encompasses a wide range of activities, from simple, localized cleanups such as variable renaming to complex, structural transformations like splitting a class or extracting a superclass, which require a comprehensive understanding of the software system. Analyzing the types of refactorings performed by agents serves as a direct indicator of their sophistication and capability. Are these agents merely acting as ``code janitors,'' automating low-level syntactic cleanup, or are they beginning to function as ``software architects,'' executing deep structural design improvements essential for long-term maintainability? By comparing the distribution of agentic refactoring types with known human refactoring patterns~\cite{icsme/horikawa25}, this RQ establishes the first empirical profile of agentic refactoring and evaluates its current level of maturity.

\subsubsection{Approach}
For each detected refactoring instance, \RMiner provides its specific type from a set of 103 distinct operations. We record these types for all \AR instances in our dataset. To understand the structural impact of these operations, we classify them using the three abstraction levels proposed by Murphy-Hill\etal~\cite{DBLP:journals/tse/Murphy-HillPB12}. Since Murphy-Hill\etal~\cite{DBLP:journals/tse/Murphy-HillPB12} did not classify all 103 refactoring types detected by \RMiner, we extended their classification framework by applying the same criteria to categorize the remaining refactoring types.  \RMiner's official benchmark evaluation\footnote{\url{https://github.com/tsantalis/RefactoringMiner/blob/master/documentation/accuracy.md}} demonstrates uniformly high precision and recall across all abstraction levels, ensuring that our results are not skewed by detection bias.  Specifically, we classified each refactoring based on whether it modifies (i) only signatures or interfaces, (ii) both signatures and code blocks, or (iii) only internal code blocks. The details of the three abstraction levels of refactoring instances are as below:

\begin{itemize}
    \item \textbf{High-level (58 types)}: Refactorings that only change the signatures of classes, methods, or fields without modifying internal code blocks. These alter the public interface and often require changes to calling code (\eg~\refactoring{Rename Method}, \refactoring{Add Parameter}, \refactoring{Move Class}, \refactoring{Change Method Access Modifier}).
    
    \item \textbf{Medium-level (21 types)}: Refactorings that change both signatures and code blocks, bridging internal logic and external structure (\eg~\refactoring{Extract Method}, \refactoring{Inline Method}, \refactoring{Move and Inline Method}, \refactoring{Change Attribute Type}).
    
    \item \textbf{Low-level (24 types)}: Refactorings confined exclusively to code blocks (typically within method bodies) that are not visible externally (\eg~\refactoring{Rename Variable}, \refactoring{Change Variable Type}, \refactoring{Extract Variable}, \refactoring{Replace Anonymous with Lambda}).
\end{itemize}

The complete mapping of all 103 refactoring types to abstraction levels is available in our replication package.\footnote{\url{https://github.com/Mont9165/Agent_Refactoring_Analysis/blob/main/Refactoring_Level_Classification.md}} We then calculate the distribution of refactoring instances across these three levels. To identify differences between agent and human behavior, we compare our results with human refactoring patterns reported in prior work~\cite{icsme/horikawa25}, mapping the human-driven instances to the same three-level framework for a direct comparison.

\begin{table}[t]
\centering
\caption{Refactoring abstraction levels: Agentic Refactoring vs Human}
\label{tab:rq2_refactoring_levels}
\begin{tabular}{lcc}
\toprule
Abstraction Level & \% Agentic Refactoring & \% Human Refactoring~\cite{icsme/horikawa25} \\ 
\midrule
High-level   & \AgenticHighPct   & \HumanHighPct \\
Medium-level & \AgenticMediumPct & \HumanMediumPct \\
Low-level    & \AgenticLowPct    & \HumanLowPct \\
\bottomrule
\end{tabular}
\end{table}

\subsubsection{Findings}
\finding{Agents and humans refactor at similar abstraction levels, with agents leaning slightly toward low-level edits.}
As shown in Table~\ref{tab:rq2_refactoring_levels}, the ranking of abstraction levels is the same for both agents and humans: high-level refactorings are the most frequent, followed by low-level and then medium-level. 
The proportions are also close. High-level refactorings account for \AgenticHighPct{} of agentic and \HumanHighPct{} of human refactorings, and medium-level refactorings for \AgenticMediumPct{} and \HumanMediumPct{}. The largest gap is in low-level refactorings (\AgenticLowPct{} vs. \HumanLowPct{}).
This small difference may arise because low-level refactorings (\eg~\refactoring{Rename Variable}, \refactoring{Change Variable Type}) need less system context than high-level ones, or because developers delegate routine cleanup to agents.

\begin{table}[t]
\centering
\caption{Comparison of Agentic and Human Refactoring Instances (Top 3 at each level)}
\label{tab:rq2_refactoring_comparison_overall}
\begin{tabular}{lllr}
\toprule
Abstraction Level & Actor & Refactoring Type & \% Instances \\
\midrule

\multirow{6}{*}{High-level}
  & \multirow{3}{*}{Agent} & \refactoring{\textbf{Move Class}}           & \AgentMoveClassPct \\
  &       & \refactoring{\textbf{Rename Attribute}} & \AgentRenameAttributePct \\
  &       & \refactoring{Rename Method}             & \AgentRenameMethodPct \\
\cmidrule{2-4}
  & \multirow{3}{*}{Human} & \refactoring{\textbf{Add Method Annotation}} & \HumanAddMethodAnnotationPct \\
  &       & \refactoring{Rename Method}             & \HumanRenameMethodPct \\
  &       & \refactoring{\textbf{Add Parameter}}    & \HumanAddParameterPct \\
\midrule
\multirow{6}{*}{Medium-level}
  & \multirow{3}{*}{Agent} & \refactoring{Change Attribute Type}          & \AgentChangeAttributeTypePct \\
  &       & \refactoring{Change Parameter Type}     & \AgentChangeParameterTypePct \\
  &       & \refactoring{\textbf{Extract Method}}   & \AgentExtractMethodPct \\
\cmidrule{2-4}
  & \multirow{3}{*}{Human} & \refactoring{Change Parameter Type}          & \HumanChangeParameterTypePct \\
  &       & \refactoring{\textbf{Change Return Type}}                     & \HumanChangeReturnTypePct \\
  &       & \refactoring{Change Attribute Type}     & \HumanChangeAttributeTypePct \\
\midrule
\multirow{6}{*}{Low-level}
  & \multirow{3}{*}{Agent} & \refactoring{Rename Variable}                & \AgentRenameVariablePct \\
  &       & \refactoring{Change Variable Type}      & \AgentChangeVariableTypePct \\
  &       & \refactoring{Rename Parameter}          & \AgentRenameParameterPct \\
\cmidrule{2-4}
  & \multirow{3}{*}{Human} & \refactoring{Change Variable Type}           & \HumanChangeVariableTypePct \\
  &       & \refactoring{Rename Variable}           & \HumanRenameVariablePct \\
  &       & \refactoring{Rename Parameter}          & \HumanRenameParameterPct \\
\bottomrule
\end{tabular}
\end{table}

\smallskip\noindent
\finding{Agents and humans share a common focus on low-level refactoring operations but diverge in high-level refactoring.} Table~\ref{tab:rq2_refactoring_comparison_overall} highlights distinct agentic priorities in the most common refactoring operations. 
At the low level, \refactoring{Rename Variable}, \refactoring{Change Variable Type}, and \refactoring{Rename Parameter} are the top three operations for both agents and humans. Together they account for \AgentTopLowSumPct{} of agentic and \HumanTopLowSumPct{} of human refactoring instances. The gap comes from \refactoring{Rename Variable} (\AgentRenameVariablePct{} vs. \HumanRenameVariablePct{}) and \refactoring{Rename Parameter} (\AgentRenameParameterPct{} vs. \HumanRenameParameterPct{}), while \refactoring{Change Variable Type} is slightly less frequent for agents (\AgentChangeVariableTypePct{} vs. \HumanChangeVariableTypePct{}). 
At the high level, however, their focus diverges. 
\refactoring{Rename Method} appears in both top-three lists (\AgentRenameMethodPct{} vs. \HumanRenameMethodPct{}), but the two most frequent agentic high-level types are \refactoring{Move Class} (\AgentMoveClassPct{}) and \refactoring{Rename Attribute} (\AgentRenameAttributePct{}), whereas \refactoring{Add Method Annotation} (\HumanAddMethodAnnotationPct{}) and \refactoring{Add Parameter} (\HumanAddParameterPct{}) appear only in the human top three. \refactoring{Move Class} is concentrated in a few commits: it appears in 136 agentic refactoring commits, and five of them contain 32\% of its instances.
At the medium level, both agents and humans frequently perform \refactoring{Change Attribute Type} and \refactoring{Change Parameter Type}. Beyond these two, agents most often apply \refactoring{Extract Method} (\AgentExtractMethodPct{}) and humans \refactoring{Change Return Type} (\HumanChangeReturnTypePct{}).

\summarybox{\textbf{Answer to RQ2}}{
High-level refactorings are the most common category for both agentic refactoring (\AgenticHighPct{}) and human refactoring (\HumanHighPct{}), and the two level distributions are close. Agentic refactoring has a slightly larger share of low-level refactorings (e.g., renaming and type changes) than human refactoring (\AgenticLowPct{} vs. \HumanLowPct{}).}

\subsection{\rqC}\label{sec:rqc}

\subsubsection{Motivation}
In human-driven development, intent is central. Developers refactor code to improve readability, reduce coupling in preparation for new features, or make the code easier to test~\cite{DBLP:journals/tse/KimZN14}. But what about agents? This research question explores the motivations behind agentic refactoring. Do developers use agents as proactive ``quality partners'' to enhance maintainability and reduce technical debt, or as tactical assistants for short-term objectives such as improving the readability of a specific method? Understanding these motivations is essential for aligning agent behavior with developer expectations.

\subsubsection{Approach}\label{sec:rqc-approach}
\begin{figure}[t]
    \centering
    \includegraphics[width=0.9\linewidth]{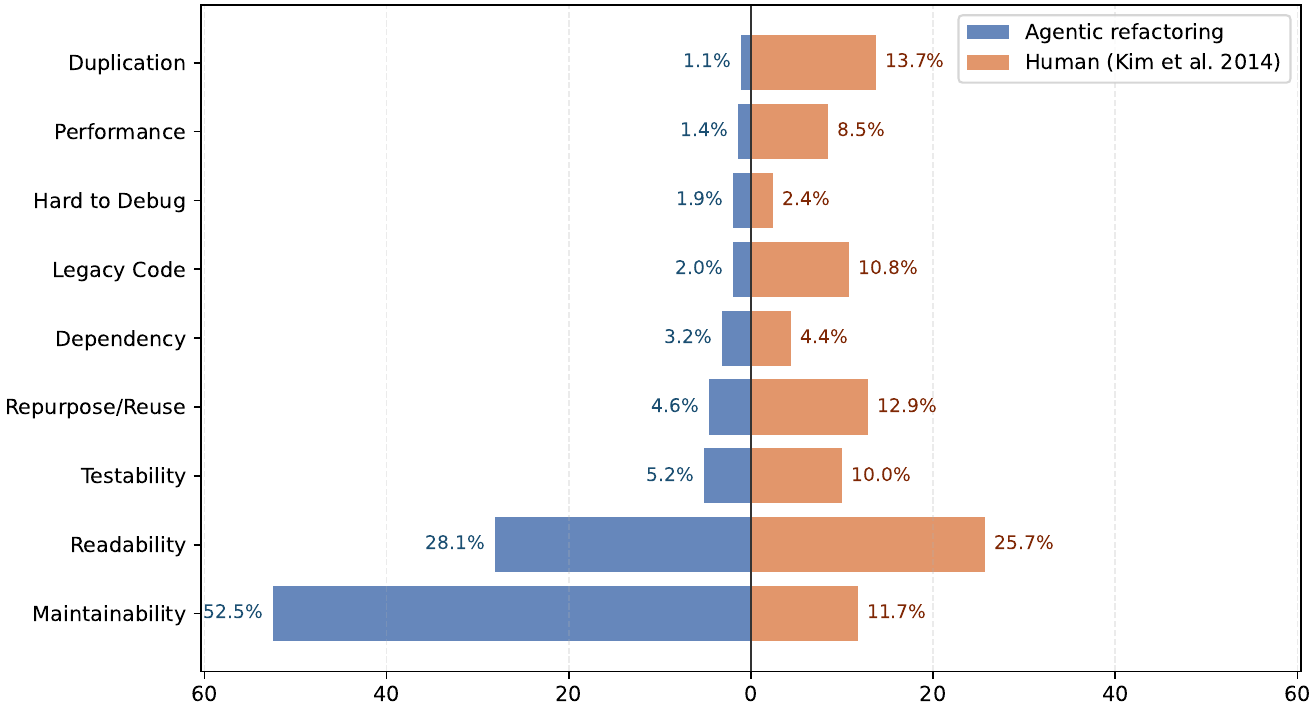}
    \caption{Refactoring purpose comparison between agents and humans~(normalized)~\cite{DBLP:journals/tse/KimZN14}}
    \label{fig:rq3_purpose}
\end{figure}

To analyze why agentic refactoring is performed, we classify the primary motivation of each \AR commit using the ten motivation categories defined by Kim\etal~\cite{DBLP:journals/tse/KimZN14} (\eg maintainability, readability, testability, legacy code\etc), which include \textit{logical mismatch}. Because \textit{logical mismatch} corresponds to bug fixing rather than a refactoring motivation, we exclude commits assigned to it and report the motivation distribution over the remaining nine refactoring-motivation categories; for consistency we also rerank the human refactoring data reported by Kim\etal~\cite{DBLP:journals/tse/KimZN14} accordingly.  In our data, GPT-4.1-mini labeled \ARLogicalMismatch{} of the \numint{\AgenticRefCommits} agentic refactoring commits (\ARLogicalMismatchPct\%) as \textit{logical mismatch}; the distribution in Figure~\ref{fig:rq3_purpose} is therefore reported over the remaining \numint{\ARMotivationBase} commits. 
The classification uses the Pull Request title, commit message, and detected refactoring types as input. 
 Due to the large sample size of the collected \AR commits, we leverage GPT-4.1-mini to automatically categorize each commit. We selected GPT-4.1-mini for its cost efficiency (see Section~4.2.2), while its classification reliability is confirmed by the human validation reported below (Cohen's $\kappa$ = \PurposeGPTKappa{}).

To validate the reliability of the automatic classification, two inspectors (each with eight years of programming experience) independently label a stratified sample of 100 \AR commits, comprising ten commits from each of the ten categories the classifier can output, including \textit{logical mismatch}. For the reliability analysis we retain the commits whose labels in the respective comparison both fall within the nine refactoring-motivation categories, excluding those where at least one label is \textit{logical mismatch} (\ie bug fixing rather than a refactoring motivation) or could not be assigned to a refactoring-motivation category; this yields \PurposeReliabilityN{} valid label pairs per comparison. We assess inter-rater reliability using Cohen's $\kappa$ coefficient~\cite{emam1999kappa}, which measures the degree of agreement between annotators.
As shown in Table~\ref{tab:rq3_reliability}, inter-rater agreement between the two human annotators is almost perfect (Cohen's $\kappa$ = \PurposeHumanKappa). Disagreements are resolved through discussion with another inspector (eighteen years of programming experience), producing the final reference labels. We then compare the GPT-4.1-mini annotations with these adjudicated human labels. Agreement between GPT-4.1-mini and the adjudicated human labels is substantial (Cohen's $\kappa$ = \PurposeGPTKappa), indicating that GPT-4.1-mini provides reliable motivation classification for large-scale empirical analysis. 

We inspected all \PurposeGPTDisagree{} disagreements among the \PurposeReliabilityN{} valid label pairs in the GPT--human comparison, and found three recurring patterns.
First, the most common confusion was between \textit{maintainability} and \textit{readability} (3 cases). These categories overlap conceptually. For example, removing unused parameters can reduce future maintenance effort and improve code clarity. GPT-4.1-mini selected maintainability, while the annotators judged readability to be the primary motivation.
Second, \textit{legacy code} accounted for the largest number of missed labels (5 cases). These cases often involved modernization commits, such as ``Update Spring Boot to 3.4.5 and fix deprecation warnings'' or ``Refactor callbacks to avoid lambdas'' for Java~7 compatibility. GPT-4.1-mini labeled the mechanism of the change, such as \textit{dependency} or \textit{repurpose/reuse}. The annotators instead labeled the underlying motivation: addressing outdated code.
Third, \textit{maintainability} often served as a broad catch-all category. It appeared in 6 of the \PurposeGPTDisagree{} GPT--human disagreements, and in 8 of the \PurposeHumanDisagree{} disagreements between the two human annotators.
These confusions occurred between closely related categories. Therefore, they do not affect the main RQ3 finding. Internal quality concerns, especially maintainability and readability, motivate most agentic refactorings. 

\begin{table}[t]
\centering
\caption{Reliability of refactoring-purpose classification (\PurposeReliabilityN{} valid label pairs per comparison)}
\label{tab:rq3_reliability}
\begin{tabular}{lrrr}
\toprule
Comparison & Cohen's $\kappa$  & Accuracy   & Macro F1  \\
\midrule
Human vs Human          & \PurposeHumanKappa & \PurposeHumanAccuracy & \PurposeHumanMacroF \\
GPT-4.1-mini vs Human   & \PurposeGPTKappa & \PurposeGPTAccuracy & \PurposeGPTMacroF \\
\bottomrule
\end{tabular}
\end{table}

\subsubsection{Findings}
\finding{Agentic refactoring is primarily driven by maintainability, a purpose that is far more dominant for agents than for humans.}
As shown in Figure~\ref{fig:rq3_purpose}, maintainability is the primary driver for more than half (\ARMaintainabilityPct) of the agentic refactoring commits with an identified refactoring motivation. This differs sharply from human refactoring patterns, where maintainability is a much less frequent purpose (11.7\%)~\cite{DBLP:journals/tse/KimZN14}. The second most common agent purpose is readability (\ARReadabilityPct), which aligns closely with the most common human purpose (25.7\%), establishing it as a shared, core priority for both. Together, these two categories account for over 80\% of all agentic refactoring purposes. 
This suggests that developers primarily dispatch agents to perform day-to-day codebase care (\eg cleanups and consistency edits) rather than to pursue broader design goals.

\smallskip\noindent
\finding{Agentic refactoring rarely targets design-level improvements, such as duplication or reuse.}
In sharp contrast to the focus on maintainability, motivations related to design-level refactoring are much less frequent in agentic commits than in human ones. For example, duplication (\ARDuplicationPct) and repurpose/reuse (\ARReusePct) are far less common motivations for agents than for humans. This is the inverse of human-driven refactoring, where these two categories are prominent (13.7\% and 12.9\%, respectively), indicating that humans frequently refactor to improve modularity and reduce redundancy. This pattern suggests that current agents focus on localized quality cleanup rather than complex, system-wide restructuring.

\smallskip\noindent
\finding{Corrective refactoring occasionally appears but remains a secondary purpose.}
A small fraction of agentic refactoring is motivated by corrective concerns. This includes addressing code that is hard to debug (\ARDebugyPct) or dealing with legacy code (\ARLagacyPct). These instances often involve tactical improvements rather than deep bug fixing, such as adding logging (\eg \texttt{log.debug()}) where it was missing,\footnote{\url{https://github.com/Braindeiko01/CRduels/commit/4ea01106f0cf74663cf1dccecd94b4475e11794a}} or modernizing code by upgrading Java versions (\eg from Java 7 to Java 17).\footnote{\url{https://github.com/BrettMiller99/ReallyOldJavaProject/commit/daea563191184899891e1a1c66e2e8d60fa94bfc}} These cases are present but clearly not the primary use case for agents.

\summarybox{\textbf{Answer to RQ3}}{
Agentic refactoring is overwhelmingly motivated by internal quality concerns. Maintainability (\ARMaintainabilityPct) and readability (\ARReadabilityPct) are the dominant drivers, accounting for over 80\% of the analyzed commits. This focus on localized cleanup contrasts with human refactoring, which more frequently addresses design-level concerns like duplication and code reuse, motivations that are much less frequent in agentic commits.
}

\subsection{\rqD}\label{sec:rqd}

\subsubsection{Motivation}
The core premise of refactoring is to improve a software system’s internal code quality. However, decades of research on human refactoring indicate that this outcome is not guaranteed. Refactorings may fail to remove existing smells or even introduce new ones~\cite{DBLP:conf/sigsoft/CedrimGMGSMFRC17, Bavota12refactorBug}, and similar uncertainty applies to agents. To justify delegating maintenance tasks to AI agents, we need quantitative evidence of their impact. This RQ shifts the focus from intent to outcomes. By analyzing established software metrics and comparing smell counts before and after agentic commits, we provide the first empirical assessment of whether agentic refactoring improves code quality. 

\subsubsection{Approach}
To quantify the impact of refactoring on internal code quality, we extract object-oriented metrics and design smells before and after each \AR commit using \designitejava~\cite{DBLP:conf/msr/Sharma24}. We select \designitejava because it is widely used in empirical software engineering research (\eg~\cite{DBLP:conf/icsm/ShahSSPSRSS23,DBLP:conf/fedcsis/LambariaC22}). As it performs static analysis directly on source code, it ensures consistent metric extraction across heterogeneous projects without requiring build configuration. The tool computes class- and method-level metrics commonly used in software quality assessment~\cite{DBLP:journals/tse/ChidamberK94}, including size (\textit{Lines of Code}), complexity (\textit{Cyclomatic Complexity}, \textit{Weighted Methods per Class}), coupling (\textit{Fan-in}, \textit{Fan-out}), cohesion (\textit{Lack of Cohesion in Methods}), and inheritance depth (\textit{Depth of Inheritance Tree}). \designitejava also detects 27 design and implementation smells (\eg \textit{Long Method}, \textit{Complex Method}, \textit{Cyclic Dependency}), enabling a multi-perspective analysis of code maintainability.

In addition, we evaluate before-and-after changes ($\Delta=\textit{after}-\textit{before}$) in internal code quality metrics detected by \designitejava for agentic refactoring commits. Before-and-after changes (\ie values of $\Delta$) are aggregated along two dimensions: (1) the abstraction level (high, medium, low) of the refactoring instances, and (2) the type of the refactoring instances. To study the differences in these code-quality metrics changes,  we perform the Wilcoxon signed-rank test~\cite{Wilcoxon1945}, a non-parametric paired test that does not assume normality, with $p$-values adjusted via the Benjamini-Hochberg FDR~\cite{GlueckMandelKarimpourFardHunterMuller+2008}. 
We also compute the rank-biserial effect size to quantify the difference. For cross-group comparisons (\eg levels or refactoring types), we use Kruskal--Wallis tests~\cite{daniel1990applied} with FDR adjustment, and when relevant, we summarize post-hoc contrasts descriptively. Given the skewed distributions of deltas, we also report the median values of these before-and-after changes.

\begin{figure}[t]
    \centering
    \begin{subfigure}{0.49\textwidth}
        \centering
        \includegraphics[width=1\linewidth]{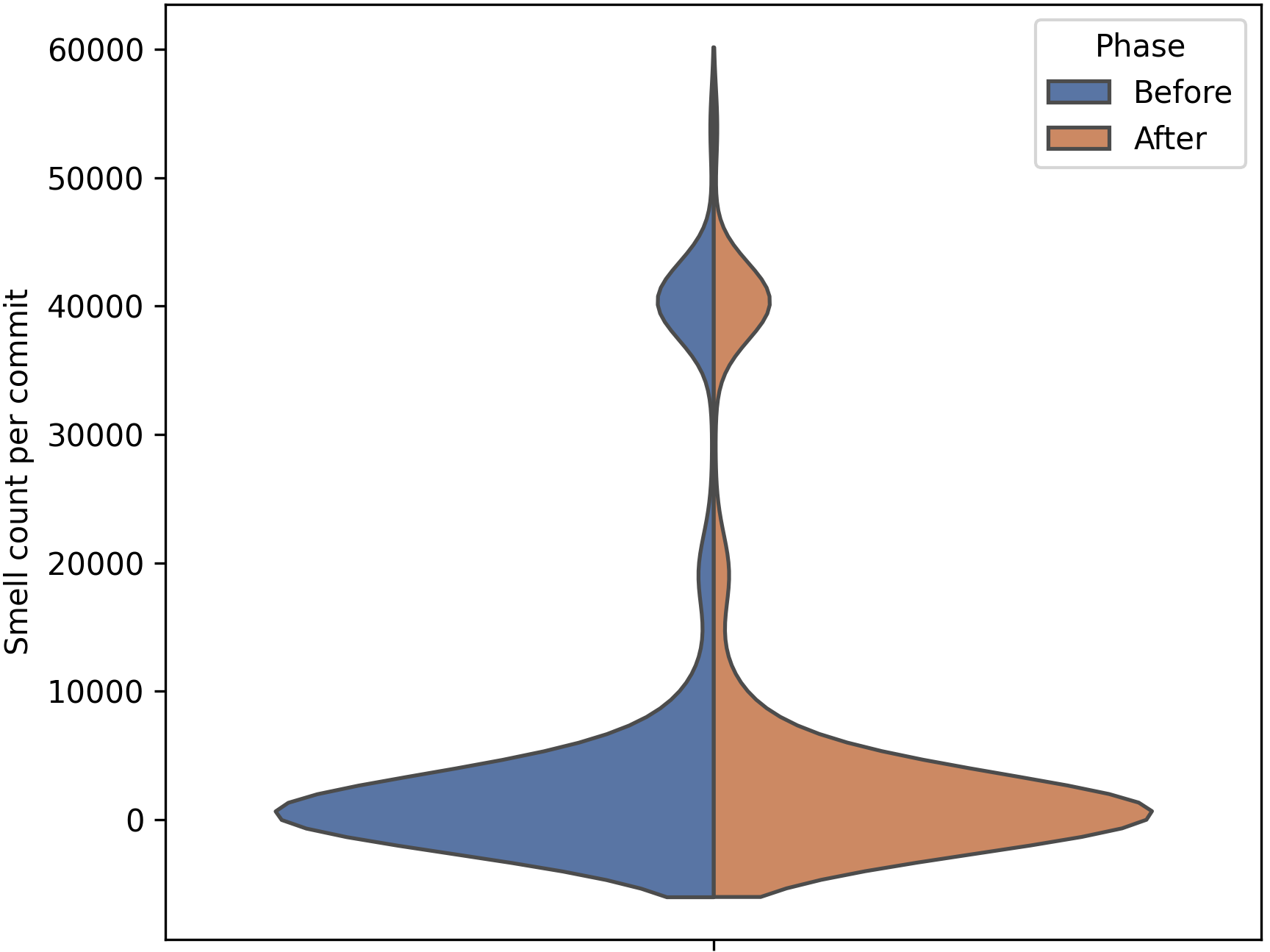}
        \caption{Implementation Smell}
        \label{fig:rq4_imp_smell}
    \end{subfigure}
    \begin{subfigure}{0.49\textwidth}
        \centering
        \includegraphics[width=1\linewidth]{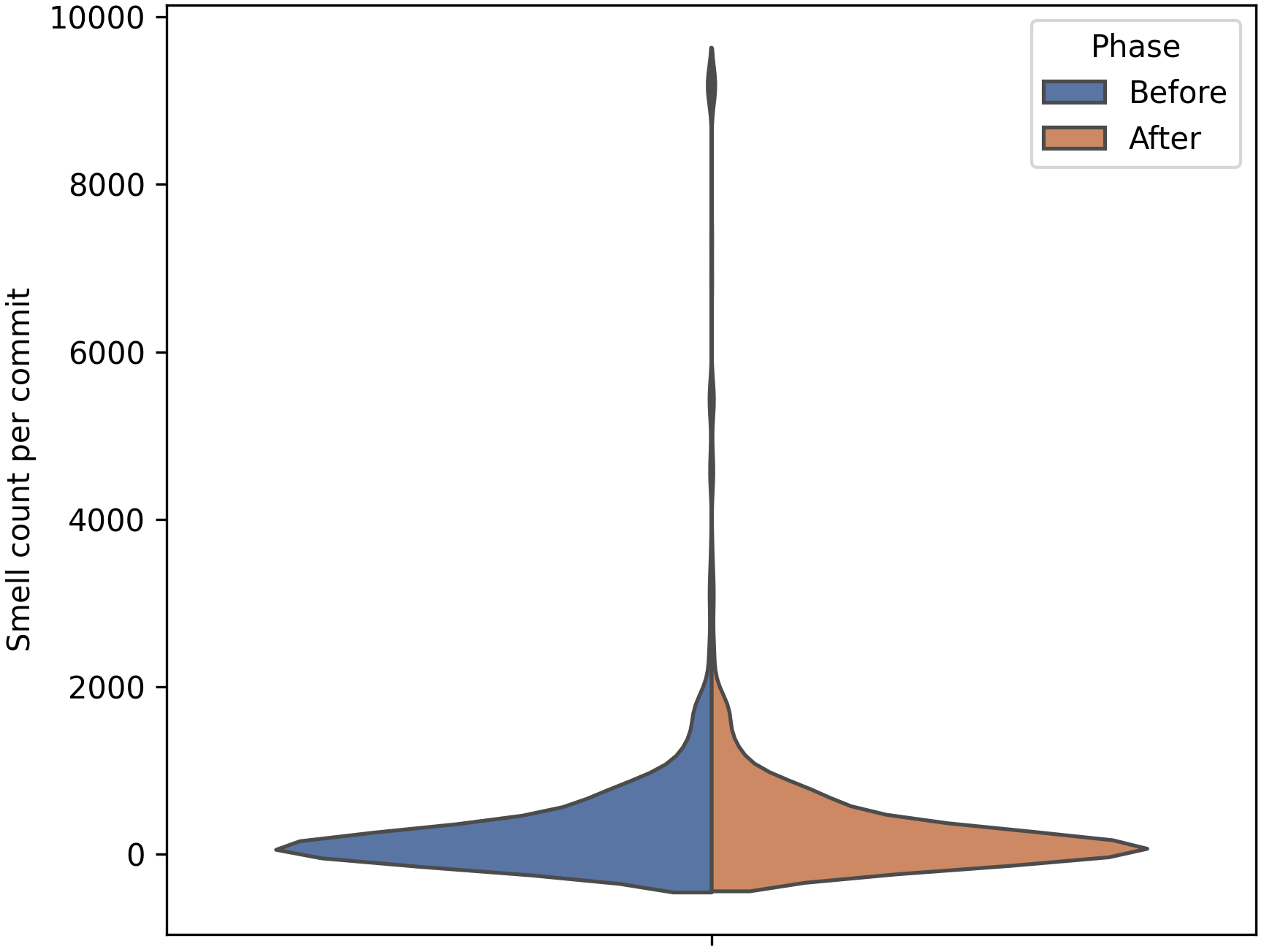}
        \caption{Design Smell}
        \label{fig:rq4_design_smell}
    \end{subfigure}
    \vspace{0.5cm}
    \caption{Smell Count Distribution (Before vs. After)}
    \label{fig:rq4_smell_distribution}
\end{figure}

\subsubsection{Findings}
\finding{Changes in design and implementation smell counts are statistically significant but practically negligible.}
Figure~\ref{fig:rq4_smell_distribution} shows that the before-and-after distributions for both design and implementation smells almost completely overlap, indicating no visible shift in typical smell levels.
Consistently, the median values remain virtually identical for both smell categories (Design: before 77.50, after 77.50; Implementation: before 355.50, after 356.00), resulting in a median $\Delta$ of 0.00 in both cases.
 Note that the median of pairwise differences can be zero even when the group-level medians differ slightly (\eg Implementation Smell before: 355.50, after: 356.00), because the median $\Delta$ is computed per commit rather than as the difference of aggregate medians. 
Although the Wilcoxon signed-rank test detects a statistically significant difference (FDR-adjusted $p \le 0.001$), the effect sizes are negligible (rank-biserial $r=-0.027$ and $-0.026$). The mean $\Delta$ values ($-6.63$ and $-32.85$ for Design and Implementation Smells, respectively) suggest that meaningful reductions occur only in a small subset of commits.
In other words, AI agents can remove smells, but they do so inconsistently, and improvements are concentrated in exceptional cases rather than being routine.

\smallskip
\noindent\finding{Agentic-refactoring commits show small improvements in structural metrics, whereas changes in smell counts are statistically detectable but not practically meaningful.}
As summarized in Table~\ref{tab:refactoring_levels_quality}, several class-level metrics improve after agentic refactoring, with the largest reductions concentrated in medium-level changes: Class LOC (median $\Delta=-15.25$), WMC (median $\Delta=-2.07$), Fan-Out (median $\Delta=-0.02$), and Fan-In (median $\Delta=-0.01$). These results indicate that agentic refactoring tends to simplify structural complexity (size and coupling) more reliably than it eliminates higher-level smell patterns.

\smallskip
\noindent\finding{Quality improvements are largest for medium-level refactorings, modest for low-level edits, and minimal for high-level signature-only changes.}
Grouping by abstraction level, medium-level refactorings produce the most consistent gains (Table~\ref{tab:refactoring_levels_quality}): 
Class LOC and WMC drop meaningfully (medians $\Delta=-15.25$ and $-2.07$), and method-level LOC decreases (median $\Delta=-1.79$). 
Low-level edits also reduce method LOC (median $\Delta=-0.42$) but show a slight increase in method cyclomatic complexity (median $\Delta=+0.08$), suggesting that these localized edits shorten methods without reducing their decision complexity. 
High-level signature-only changes show near-zero medians across most metrics, which is consistent with interface adjustments rather than structural simplification.
Kruskal--Wallis tests across levels (FDR-adjusted) support these differences at $\alpha=0.05$.

\begin{table}[!t]
\centering
\small
\caption{Median $\Delta$ (after–before) of structural metrics across refactoring abstraction levels in agentic refactoring commits. Negative values indicate improvement.}
\label{tab:refactoring_levels_quality}
\begin{tabular}{lrrr}
\toprule
Metric & Low-level & Medium-level & High-level \\
& (code block) & (signature + block) & (signature) \\

\midrule
Class-Level – Depth of Inheritance Tree & -- & 0.00 & 0.01 \\
Class-Level – Fan-Out & -- & -0.02 & 0.00 \\
Class-Level – Fan-In & -- & -0.01 & 0.00 \\
Class-Level – Number of Methods & -- & -0.10 & 0.01 \\
Class-Level – Weighted Methods per Class & -- & -2.07 & 0.03 \\
Class-Level – Lines of Code & -- & -15.25 & 0.10 \\
Method-Level – Parameter Count & 0.00 & 0.01 & 0.11 \\
Method-Level – Cyclomatic Complexity & 0.08 & 0.01 & 0.00 \\
Method-Level – Lines of Code & -0.42 & -1.79 & 0.11 \\
\bottomrule
\multicolumn{4}{l}{-- Not applicable: Class-Level metrics are not measured for Low-level (code block) refactorings.}
\end{tabular}
\end{table}

\smallskip
\noindent\finding{Structural refactorings (\eg decomposition and modularization) show the largest metric reductions.}
Analyzing median metric deltas by refactoring type reveals that refactorings introducing new structure or decomposing responsibilities have the largest metric reductions. For example, \refactoring{Extract Subclass} yields large median reductions in Class LOC ($\Delta=-87.5$) and WMC ($\Delta=-11.5$), while \refactoring{Split Class} similarly reduces Class LOC ($\Delta=-16.0$) and WMC ($\Delta=-4.0$).\footnote{\url{https://github.com/Mont9165/Agent_Refactoring_Analysis/blob/main/outputs/designite/group_heatmaps/refactoring_heatmap_sar_median_delta.csv}}
{These refactorings improve modularity and reduce class-level complexity by distributing responsibilities across more cohesive components. However, the metrics reported in this study are measured at the class level rather than the project level. While extraction-based refactorings reduce metrics for the refactored class (\eg~\refactoring{Extract Subclass} yields Class LOC $\Delta$ = $-$87.5 and WMC $\Delta$ = $-$11.5), they necessarily create new classes that inherit part of the original complexity. Moreover, distributing logic across more classes can increase inter-class coupling (\eg Fan-Out), even as intra-class complexity decreases. Fowler~\cite{DBLP:books/daglib/0019908} argues that distributing responsibilities across smaller, cohesive classes improves maintainability because each unit becomes independently comprehensible and testable, even if aggregate complexity does not decrease. Nevertheless, without project-level aggregation of metrics (\eg summing WMC or LOC across all classes before and after the commit), our results should be interpreted as improvements to the refactored entity rather than definitive evidence of net project-level quality gains.}
In contrast, refactorings that preserve overall structure, such as signature adjustments or local rewrites, tend to produce smaller metric shifts.
These findings indicate that the impact of agentic refactoring is strongly influenced by refactoring type. While structural decomposition refactorings drive measurable improvements, localized edits have a more limited effect on structural indicators.

\smallskip
\noindent\finding{Not all refactoring types improve measured metrics; some primarily support clarity or evolution.}
Several high-frequency agentic refactoring types (\eg identifier renames and type changes) show negligible before-and-after change in the analyzed structural metrics after FDR correction. 
This does not imply that these refactorings lack value; rather, their main benefits (\eg readability, naming consistency, API clarity) are not captured by the selected design-level indicators. 
By contrast, transformation types with both semantic and structural effects may yield mixed outcomes. 
For example, \refactoring{Move And Inline Method} tends to reduce method LOC (median $\Delta=-0.5$) but can leave cyclomatic complexity unchanged, or even increase it slightly when logic is redistributed across helper methods. 
These observations emphasize that metric-based evaluation should be interpreted together with developer intent (see \sec{sec:rqc}). Agentic refactoring is often used for code consistency and comprehension.

\summarybox{\textbf{Answer to RQ4}}{
Agentic refactoring yields statistically significant but generally small structural improvements, most notably for medium-level changes that reduce class size and complexity (\eg Class LOC median $\Delta=-15.25$, WMC median $\Delta=-2.07$). 
Design and implementation smell counts exhibit FDR-significant but practically negligible changes (median $\Delta=0.00$ for both; rank-biserial $r=-0.027$ and $-0.026$), with larger reductions concentrated in a small subset of commits.
Refactoring type matters: structural decomposition refactorings (\eg~\refactoring{Extract Subclass}, \refactoring{Split Class}) produce clearer metric gains, whereas renames and other low-level edits rarely affect structural indicators.
}

\section{Implications}\label{sec:implications}

In this section, we discuss the implications of our findings for researchers, developers, and coding agent builders.

\subsection{Implications for Researchers}
\noindent\textbf{Researchers should investigate the hidden burden of implicit refactoring on development workflows.}
\sec{sec:rqa} shows that more than a third of refactoring instances (\NonSARInstancesPct\%) occur in commits without explicit refactoring intent, which may indicate tangled changes. This suggests that agents frequently perform refactorings as a side effect of other tasks. Such tangling can increase review effort, as developers must validate both the primary task (\eg a feature) and the incidental refactorings to ensure behavior is preserved. Prior work indicates that developers want to verify logic preservation~\cite{DBLP:journals/infsof/SergeyukGBA25}, so a high volume of implicit refactoring could lengthen reviews and reduce trust. Future work should quantify this burden by analyzing review comments, merge latency, and revert rates for commits with implicit refactoring.

\smallskip
\noindent\textbf{Researchers should create benchmarks for high-level refactoring.}
\sec{sec:rqb} reports how often agents perform each type of refactoring, but not whether they perform it correctly. Assessing this requires curated benchmarks and gold standards for architectural refactorings (\eg~\refactoring{Extract Class}, \refactoring{Introduce Parameter Object}, \refactoring{Move Class}). These refactorings change interfaces and require coordinated updates to callers, so correctness is harder to judge from frequency alone. Datasets should include ground-truth intent, behavior-preserving tests, and expected post-conditions (reduced WMC/fan-out).

\smallskip
\noindent\textbf{Researchers should study the disconnect between agentic refactoring and measured quality outcomes.}
The majority~(86.9\%) of agentic PRs are merged into the software projects~(\sec{subsec:dataset_overview}), and our findings reveal that 26.1\% of agentic commits contain explicitly stated refactoring (\sec{sec:rqa}). However, traditional quality metrics show minimal improvement (\sec{sec:rqd}). Researchers should investigate this gap to understand whether these agentic refactorings (\eg renaming variables) provide tangible benefits to human developers, such as improved comprehension or reduced cognitive load, or whether agents are engaging in low-value ``code churn'' that fails to address deeper structural issues. 
 In particular, our findings reveal that while \ARMaintainabilityPct~of agentic refactoring is motivated by maintainability (Section 5.3), the frequently occurring low-level operations primarily improve naming consistency and readability rather than structural metrics (Section 5.4). This suggests that current agents interpret ``maintainability'' narrowly as code-level cleanup rather than structural simplification. The benefits of these refactorings (\eg reduced cognitive load, improved naming clarity) may be real but are not captured by traditional design-level indicators such as WMC or Fan-Out.

\smallskip
\noindent\textbf{Researchers should validate long-term quality outcomes through longitudinal analysis for agentic refactorings.}
Our before-and-after analysis of structural metrics~(\sec{sec:rqd}) shows small but statistically significant improvements, particularly for medium-level changes (\eg Class LOC median $\Delta=-15.25$, WMC median $\Delta=-2.07$). The core premise of refactoring, however, concerns long-term evolvability and maintainability. Although agentic refactoring is overwhelmingly motivated by maintainability~(\sec{sec:rqc}), it does not consistently reduce design and implementation smells. Researchers should assess whether these structural gains translate into long-term benefits by conducting longitudinal studies that track defect density, post-release defects, and effort for future modifications.

\subsection{Implications for Developers}

\noindent\textbf{Developers should demand commit hygiene from coding agents so that refactoring does not silently accompany behavioral changes.}
More than a third of all detected refactoring instances (\NonSARInstancesPct\%) appear in commits whose messages do not mention refactoring (\sec{sec:rqa}), indicating that agents frequently mix renames and restructurings with feature or bug-fix changes. Such tangled changes bias code review and downstream analyses~\cite{DBLP:journals/ese/HerzigJZ16} and make behavior preservation harder to verify. Developers should therefore (i)~instruct agents through agent context files~\cite{DBLP:journals/corr/abs-2511-12884} to avoid out-of-scope refactoring, (ii)~require cleanup changes to be isolated in separate commits or PRs, and (iii)~highlight mixed-intent changes during review, potentially using automatic changeset decomposition~\cite{DBLP:conf/icse/BarnettBBL15}.

\smallskip
\noindent\textbf{Developers should use coding agents for routine cleanup tasks.}
\sec{sec:rqb} shows that agentic refactoring contains a slightly larger share of low-level refactorings than human refactoring (\AgenticLowPct{} vs. \HumanLowPct{}), and that its two most common operations are \refactoring{Rename Variable} (\AgentRenameVariablePct{}) and \refactoring{Change Variable Type} (\AgentChangeVariableTypePct{}). These findings suggest that developers already use current agents for localized, repetitive maintenance tasks, including naming cleanup, type modernization, and consistency updates. 
Prior studies have shown that software maintenance often involves recurring change patterns that can be automated~\cite{DBLP:conf/icse/DilharaKSD22, DBLP:conf/kbse/NguyenNNNR13}. Coding agents may therefore reduce the manual effort required to apply these changes across a repository. For example, agents can conduct periodic repository-wide consistency sweeps, allowing developers to focus on reviewing and validating the resulting changes.

\smallskip
\noindent\textbf{Developers may obtain the greatest value from coding agents by identifying architectural hotspots and assigning the restructuring to agents.}
Not all refactoring types are equally effective~(\sec{sec:rqd}). Structural refactorings (such as \refactoring{Extract Subclass} and \refactoring{Split Class}) produce larger reductions in class size and complexity than common low-level edits. This finding suggests an effective collaborative workflow in which developers identify maintainability problems and agents carry out the mechanical restructuring. In this workflow, developers provide the diagnosis, while agents perform the implementation.

\subsection{Implications for Coding Agent Builders}

\smallskip
\noindent\textbf{Coding agent builders should equip agents with specialized tools to autonomously detect and fix design flaws.}
A key finding in \sec{sec:rqd} is that agentic refactoring fails to consistently reduce the overall count of design and implementation smells (median $\Delta$ of 0.00). This suggests agents are overlooking specific flaws. Prior work on RepairAgent~\cite{DBLP:conf/icse/BouzeniaDP25} demonstrates that agents equipped with task-specific tools can autonomously perform automated program repair by gathering information, applying patches, and validating candidate fixes. Similarly, agentic refactoring could be improved by equipping agents with refactoring-specific analysis tools (\eg~\designitejava) and feeding their findings back into the planning loop. This setup would allow the agent to actively seek out smells, verify improvements, and autonomously decide when to run a smell scan, retrieve metrics, and re-invoke transformations.

\smallskip
\noindent\textbf{Coding agent builders should generate the evidence reviewers need to assess both safety and value.}
A generated refactoring should arrive with evidence for two distinct claims: that it preserves observable behavior, and that it advances the stated quality objective. For the former, agents should report the affected APIs and callers, the tests, builds, and differential checks that were run, and the scope and limitations of those checks. For the latter, agents should report how the indicators that match the stated objective changed; for readability-oriented changes, this evidence should not rely on structural metrics alone, which can remain unchanged even when clarity improves~(\sec{sec:rqd}). When mixed changes cannot be avoided, the review interface should group behavior-changing edits separately from restructuring. Industrial deployments of automatic program repair follow the same approach: generated patches are integrated into existing CI and review workflows~\cite{marginean2019sapfix, urli2018repairnator, williams2024bloomberg}, and practitioners assess them more readily when test reports, coverage information, and patch explanations accompany them~\cite{noller2022trust, winter2023developers}.

\section{Threats to Validity}\label{sec:threats_to_validity}

In this section, we discuss the threats to validity of our study about agentic refactoring.

\subsection{Internal Validity}
Threats to internal validity concern potential confounding factors that could influence our results. First, our study relies on automated tools for data analysis. \RMiner\cite{DBLP:journals/tse/TsantalisKD22,DBLP:journals/tosem/AlikhanifardT25}, which we used to identify refactoring operations, is known to produce both false positives and false negatives.
We classified \RMiner's 103 refactoring types into three abstraction levels based on their structural impact (Section~5.2.2): the catalog contains 58 high-level, 21 medium-level, and 24 low-level types, indicating that the catalog is not biased toward low-level operations. Furthermore, \RMiner's benchmark evaluation\footnote{\url{https://github.com/tsantalis/RefactoringMiner/blob/master/documentation/accuracy.md}} shows uniformly high precision and recall across all abstraction levels, confirming that detection accuracy does not vary substantially between levels. The same tool is applied to both agentic and human refactoring data~\cite{icsme/horikawa25}, ensuring that any residual detection characteristic affects both groups. 
Similarly, \designitejava~\cite{DBLP:conf/msr/Sharma24}, used for calculating code quality metrics, has its own inherent limitations. Although these tools are standard in empirical software engineering, their potential inaccuracies could have affected our quantitative results.
Second, for RQ3, we employed GPT-4.1-mini to automatically classify the purpose of each refactoring. To mitigate the risk of misclassification, two authors manually labeled a stratified sample of 100 refactoring commits (\PurposeReliabilityN{} valid label pairs). Agreement between the two annotators was almost perfect (Cohen's $\kappa$ = \PurposeHumanKappa{}), and agreement between GPT-4.1-mini and the adjudicated human labels was substantial (Cohen's $\kappa$ = \PurposeGPTKappa{}). This suggests that while the automated classification may not be perfect, the resulting distribution of purposes is largely reliable.
Third, our identification of refactoring intent relies on keyword matching in commit messages (Section~4.2.4). This approach may miss commits with unstated refactoring intent and may include false positives from broad patterns (\eg ``fix*'' matching bug-fix commits that incidentally contain refactoring keywords). We mitigate this risk by requiring both keyword match and \RMiner detection, but acknowledge that intent classification remains approximate. Consistent with this, GPT-4.1-mini judged \ARLogicalMismatch{} of the \numint{\AgenticRefCommits} agentic refactoring commits (\ARLogicalMismatchPct\%) to be motivated by bug fixing rather than refactoring; these are excluded from the RQ3 motivation distribution. 

\subsection{Construct Validity} 
Construct validity threats relate to the alignment between our theoretical constructs and our measurements. {First, our code quality analysis in \rqd{} measures metrics at the class level using \designitejava{}. Extraction-based refactorings (\eg~\refactoring{Extract Subclass}, \refactoring{Split Class}) reduce metrics for the refactored class but may redistribute rather than eliminate complexity at the project level. A project-level evaluation that aggregates metrics across all affected classes before and after each commit would be needed to fully assess the net impact on overall system quality. We leave this for future work. }

 Additionally, our analysis focuses on commits as the unit of analysis, as dictated by \RMiner's requirement to compare consecutive revisions. While we classify refactoring purpose (RQ3) using the PR title, commit message, and detected refactoring types, richer development artifacts such as PR discussions, code review comments, and issue context are not systematically analyzed. These artifacts could reveal, for example, whether a refactoring was initiated by the developer or autonomously by the agent. For RQ1, RQ2, and RQ4, the relevant data are the code changes and metrics themselves, which are not influenced by these artifacts. Nevertheless, qualitative analysis of PR discussions represents an important avenue for future research. 

A primary concern is the definition of an agentic commit. 
We identify such commits based on keywords and author information from the commit history. However, it is challenging to ascertain the precise extent of human intervention; developers may modify, accept, or reject parts of AI-generated code before committing. To acknowledge this ambiguity, we explicitly frame our study as an analysis of human-AI collaborative refactoring rather than purely autonomous AI contributions.

\subsection{External Validity} 
Threats to external validity concern the generalizability of our findings. 
Our study is based on the AIDev dataset \cite{DBLP:journals/corr/abs-2507-15003}, which involves a two-fold selection: it covers (i)~only publicly available open-source software (OSS) repositories on GitHub and (ii)~among those, only projects in which AI coding agents had already been adopted at the time of data collection. Although our commit collection covers repositories regardless of star count (see Section~\ref{subsec:data_mining}), our findings therefore characterize the current refactoring practice of early adopters in public OSS. Proprietary or enterprise codebases, whose development practices, coding standards, and review processes may differ significantly, are not represented.

Furthermore, our analysis was limited to Java due to the maturity of \RMiner (F-score of 99.5\%) and \designitejava for code quality metrics, as well as the need for direct comparison 
with prior human refactoring results~\cite{icsme/horikawa25}. While multi-language refactoring detectors exist (\eg RefDiff~\cite{DBLP:journals/tse/SilvaSSTV21}), they support fewer refactoring types. The prevalence and characteristics of agentic refactoring may differ in languages with distinct ecosystems (\eg Python, TypeScript). The same caution applies to languages that are co-executable with Java on the JVM (\eg Kotlin, Scala, and Groovy): although they interoperate with Java bytecode, their idioms, such as Kotlin's data classes and null-safety constructs or Scala's functional style, change both the need for and the shape of common refactorings, and refactoring-detection tools with accuracy comparable to \RMiner{} are not yet available for them. Our findings thus apply most directly to Java, and extending this study across languages, including JVM languages, is an important direction for future work. 
Therefore, caution should be exercised when generalizing our results to other contexts.

\section{Conclusion}\label{sec:conclusion}
This study provides the first large-scale empirical analysis of refactoring in agentic software development, examining \numint{\AgenticRefInstances} refactoring instances produced in agentic coding workflows across real-world open-source Java projects. Our findings clarify the current capabilities, typical uses, and impacts of agentic refactoring in open-source Java projects that have adopted AI coding agents. 

Our empirical results show that refactoring is a common and intentional activity for AI agents, explicitly targeted in \AgentRefCommitPct\% of agentic commits. This demonstrates that agents actively participate in software maintenance, frequently engaging in restructuring activities beyond feature implementation or bug fixing. 
 The stated motivations are overwhelmingly focused on internal code quality: maintainability (\ARMaintainabilityPct) and readability (\ARReadabilityPct) account for over 80\% of cases. 

Although high-level refactorings are the largest abstraction-level category overall, the two most frequent individual agentic operations are low-level, consistency-oriented edits (\refactoring{Rename Variable} and \refactoring{Change Variable Type}). Compared to human refactoring, agentic refactoring has a slightly smaller share of high-level refactorings (\AgenticHighPct{} vs. \HumanHighPct{}) and a slightly larger share of low-level ones (\AgenticLowPct{} vs. \HumanLowPct{}).

Our quantitative assessment shows that agentic refactoring produces statistically significant but small structural improvements, particularly for medium-level changes that combine signature and block modifications. These include measurable reductions in Class Lines of Code (Class LOC median $\Delta = -15.25$) and Weighted Methods per Class (WMC median $\Delta = -2.07$). However, despite the goal of improving quality, agents currently fail to consistently reduce the overall count of known design and implementation smells.

In conclusion, agentic coding tools serve as cleanup partners in open-source Java projects, frequently used for localized refactoring and consistency improvements that support long-term maintainability. Future work can extend this evaluation to other languages, ecosystems, and proprietary settings. Moreover, to realize the vision of agents as ``software architects,'' significant advancements are needed to enable autonomous, architecturally-aware restructuring that consistently addresses higher-level design smells.

\begin{acks}
We gratefully acknowledge the financial support of JSPS KAKENHI grants (JP24K02921, JP25K03100, JP26H02500), as well as JST BOOST grant (JPMJBS2423), JST ACT-X grant (JPMJAX26MH), JST ASPIRE grant (JPMJAP2415), JST CREST grants (JPMJCR23M1, JPMJCR26X7), and JST FOREST grant (JPMJFR252W). We also acknowledge the support of the Natural Sciences and Engineering Research Council of Canada (NSERC).
\end{acks}


\balance
\bibliographystyle{ACM-Reference-Format}
\bibliography{references}


\end{document}